# Numerical Fréchet derivatives of the displacement tensor for 2.5-D frequency-domain seismic full-waveform inversion in viscoelastic TTI media

Qingjie Yang*, Waves and Machine Intelligence team, Directed Energy Research Centre, Technology Innovation Institute, Abu Dhabi P.O. Box 9639, UAE, qingjie.yang@tii.ae

Bing Zhou, Department of Earth Science, Khalifa University of Science and Technology, Abu Dhabi, UAE, bing.zhou@ku.ac.ae

Marcus Engsig, Waves and Machine Intelligence team, Directed Energy Research Centre, Technology Innovation Institute, Abu Dhabi P.O. Box 9639, UAE, marcus.engsig@tii.ae

Mohamed Kamel Riahi, Department of Applied Mathematics, Khalifa University of Science and Technology, Abu Dhabi, UAE, riahi.mk@gmail.com

Mohammad Al-khaleel, Department of Applied Mathematics, Khalifa University of Science and Technology, Abu Dhabi, UAE and Yarmouk University, Department of Mathematics, Irbid 21163, Jordan, mohammad.alkhaleel@ku.ac.ae

Stewart Greenhalgh, Department of Earth Sciences, Swiss Federal Institute of Technology (ETH) Zurich, Institute of Geophysics, Zurich, Switzerland, gstewart@retired.ethz.ch

## ABSTRACT

Derivatives of the displacement tensor with respect to the independent model parameters of the subsurface, also called Fréchet derivatives (or sensitivity kernels), are a key ingredient for seismic full-waveform inversion with a local-search optimization algorithm. They provide a quantitative measure of the expected changes in the seismograms due to perturbations of the subsurface model parameters for a given survey geometry. Since 2.5-D wavefield modeling involves a real point source in a 2-D geological model with 3D (spherical) wave properties, it yields synthetic data much closer to the actual practical field data than the commonly used 2-D wave simulation does, which uses an unrealistic line source in which the waves spread cylindrically. Based on our recently developed general 2.5-D wavefield modeling scheme, we apply the perturbation method to obtain explicit analytic expressions for the derivatives of the displacement tensor for 2.5-D/2-D frequency-domain seismic full-waveform inversion in general viscoelastic anisotropic media. We then demonstrate the numerical calculations of all these derivatives in two common cases: (i) viscoelastic isotropic and (ii) viscoelastic tilted transversely isotropic (TTI) solids. Examples of the differing sensitivity patterns for the various derivatives are investigated and compared for four different homogeneous models involving 2-D and 2.5-D modeling. Also, the numerical results are verified against the analytic solutions for homogeneous models. We further validate the numerical derivatives in a heterogeneous viscoelastic TTI case by conducting a synthetic data experiment of frequency-domain full-waveform inversion to individually recover the twelve independent model parameters (density, dip angle, five elastic moduli, and five corresponding Q-factors) in a simple model comprising an anomalous square box target embedded in a uniform background. Another multi-target model experiment presenting impacts from four common seismic surveying geometries validates the Fréchet derivatives again.

# INTRODUCTION

Full-waveform inversion (FWI) is a challenging data fitting procedure that seeks to match the synthetically generated data for a progressively updated model of the geological medium with the actual observed data. It uses all the available dynamic information in the seismograms (amplitudes, travel times, frequencies) to obtain high-resolution images of the physical properties of the subsurface. It has been applied to a variety of model types, including viscoacoustic media (Gauthier et al., 1986; Keating and Innanen, 2019; Operto et al., 2004), viscoelastic media (Brossier, 2011; Brossier et al., 2009; Mora, 1987; Sun et al., 2017), and even poroelastic media (Barros et al., 2010; Yang and Malcolm, 2021), by iteratively minimizing the misfit between the simulated and observed seismograms (Virieux et al., 2017; Virieux and Operto, 2009). Although FWI was originally proposed in the time domain (Tarantola, 1984; Mora, 1987), it has also been developed in the frequency domain (Pratt et al., 1998; Shin et al., 2007; Sirgue and Pratt, 2004; Zhou and Greenhalgh, 2003), as well as in the Laplace-domain (Ha and Shin, 2012; Shin and Ho Cha, 2009). Frequency-domain FWI offers the advantages of performing stable multiscale inversion (starting at low frequencies and progressively increasing to higher frequencies) and improve modeling efficiency for multiple sources, because of two attractive features: (i) a single decomposition of the stiffness (modulus) matrix may efficiently provide the wavefield solutions at a given frequency for all sources and receivers, and (ii) a certain range of frequency components of the observed seismograms can be selected as the input data to reconstruct the subsurface geological structures at different scales. Moreover, with the rapid development of computer technology, the implementation of seismic FWI has been extended from 2-D to 3-D (Bozdağ et al., 2016; Tape et al., 2009; Vigh and Starr, 2008; Zhu et al., 2012) and from reconstructing a single model parameter (e.g. P-wave velocity) to multiple model parameters simultaneously, e.g., P-wave velocity, S-wave velocity, density (Operto et al., 2013; Operto et al., 2015; Prieux et al., 2013; Tournier et al. 2021). However, 3-D FWI is still quite expensive not only for the data acquisition but also in terms of computer cost (memory storage and runtime). Multi-parameter FWI presents a challenge and requires a novel scheme to overcome the drawback of trade-offs or crosstalk between the different parameters (Keating and Innanen, 2019; Sun et al., 2017; Yang and Malcolm, 2021).

Seismic attenuation, expressed in terms of the dimensionless inverse quality factor $1/Q$, is a key model parameter that significantly influences wavefields. It has been considered not only in the forward wavefield modeling (Kurzmann et al., 2013) but also in FWI (Fichtner and van Driel, 2014; Hicks and Pratt, 2001; Kamei and Pratt, 2013; Romanowicz, 1995; Tromp et al., 2005). Recently, the imaging of high-resolution Q has been incorporated into the framework of FWI and has attracted much attention (Bai and Tsvankin, 2019; Operto and Miniussi, 2018; Wang et al., 2019).

Most FWI strategies require the calculation of the sensitivity kernels for updating the model parameters from the current model, as the iterations proceed. These so-called Fréchet derivatives are the numerical derivatives of the displacement tensor with respect to the subsurface model parameters. For example, Tromp et al. (2005) presented the general time-domain formulations of the derivatives in elastic anisotropic media by the adjoint method. Zhou and Greenhalgh (2009) devised a perturbation method to obtain explicit analytic expressions of the 3-D frequency-domain derivatives for anisotropic elastic media with respect to the elastic moduli and density. Fichtner and van Driel (2014) adapted the adjoint method to compute the time-domain derivatives for both frequency-dependent and frequency-independent Q-factor. Yang et al. (2016) developed a time-domain systematic formulation of 3-D multi-parameter FWI based on the generalized Maxwell/Zener viscoelastic model. We also refer the reader to Wang et al. (2019) who demonstrated a time-domain Q-factor FWI scheme based on the simplified viscoacoustic equation that applies a standard linear solid model to describe the attenuation behavior. However, it is well known that the attenuation property of a medium, as defined by complex-valued moduli in the frequency-domain, is much easier to compute than in the time-domain using spring-dashpot combination mechanical models, such as the generalized Maxwell or Zener models (Carcione, 2007; Hicks and Pratt, 2001; Liu and Greenhalgh, 2019; Moczo and Kristek, 2005).

The frequency-dependent Q-factors can be employed to define the seismic wave attenuation in arbitrary viscoelastic anisotropic models. Therefore, the model parameterization includes not only the elastic moduli, but also the Q-factors, all of which are functions of spatial location. Such a multi-parameterization leads to multi-parameter FWI (Fabien et al., 2017a, 2017b). The crosstalk issue often arises in multi-parameter FWI whereby there can be tradeoffs between the different parameters, making it difficult to separate them and can even produce severe errors in the reconstruction. By separating the inversion parameters into two classes (the velocity parameter class and the attenuation parameter class), Kamei and Pratt (2013) propose frequency-domain FWI strategies for viscoacoustic media. Operto and Miniussi (2018) study a real-data case to show the feasibility of the multi-parameter frequency-domain FWI. They successfully reconstruct the vertical wave speed, density, and Q-factor distributions using multi-source stationary-receiver wide azimuth ocean-bottom cable data.

Simple 2-D surface seismic profiling is often able to provide satisfactory images of the underground structure effectively and efficiently, especially in areas where the geology is 2D or the seismic profile can be run perpendicular to the geological strike direction, such 2D experiments are also useful and relevant with vertical seismic profiling surveys (VSP's) and cross-hole seismic surveys between two wells. For these situations, the implementation of a 2-D FWI is more practical and economical than carrying out a 3-D FWI. Moreover, there would be far too many model parameters to account for or resolve in 3D with such a limited survey geometry. However, most 2-D elastic or 2-D viscoelastic FWI schemes are theoretically based on pure 2-D wavefield modeling that uses an unrealistic line-source, as mentioned previously. This necessitates the conversion of the observed point-source field data to line-source data before the 2-D FWI can be performed. The data conversion is often accomplished by a simple filter, which is derived by an analytical comparison between the asymptotic Green's functions of a point-source (3-D) and a line-source (2-D) in a homogeneous acoustic full space. Williamson and Pratt (1995) and Auer et al. (2013) found that this simple filter may introduce significant errors when the 2-D geological model is heterogeneous. Therefore, rigorous 2.5-D wave modeling becomes necessary to either implement a 2.5-D seismic FWI (Zhou et al., 2012) or to investigate the exact conversion between the line-source data and point-source data for complex geological models (Forbriger et al., 2014; Schäfer et al., 2014; Zhou et al., 2019; Yang et al., 2022). In particular, Zhao et al. (2017) and Yang et al. (2020) demonstrated the finite-element method and the Gaussian-quadrature-grid approximations to simulate both 2-D and 2.5-D frequency-domain seismic wavefields in viscoelastic anisotropic media (VEAM).

It is clear that 2.5-D numerical forward modeling methods are required for 2.5-D frequency-domain seismic FWI in complex models such as VEAM, because they generate accurate 2.5-D wavefields for computing the derivatives of the displacement tensor with respect to all the model parameters. These Fréchet derivatives are the key ingredient of the local search minimization method used in 2.5-D FWI. It has been shown that the 2-D and 3-D wave equations give the linear and positive self-adjoint differential operators, one can directly apply the adjoint method (see example, Zhou and Greenhalgh, 2009) for the explicit Fréchet derivatives. However, it has been shown that the 2.5-D wave equation gives a combination of positive and negtive self-adjoint differential operators (Zhou and Greenhalgh, 2011), we have to employ a more general pertubation method than the common adjoint method to deal with the 2.5-D case. To the best of our knowledge, the 2.5-D Fréchet derivatives of the displacement vector in a VEAM model have not been reported in previous literatures In this paper, based on our previous theoretical results of the general perturbation method for the 2.5-D elastic anisotropic media (Zhou and Greenhalgh, 2011) and our newly-developed 2.5-D viscoelastic anisotropic wave modeling (Yang et al., 2020), we derive the explicit analytic formulae of the derivatives in a general VEAM model. We discretize the model by flexible blocks, and then apply our newly-developed 2.5-D wave modeling method to obtain numerically the Fréchet derivatives of the displacement vector with respect to all the independent moduli and Q-factors in a common viscoelastic anisotropic medium called a tilted transversely isotropic (TTI) solid. To verify the applicability of the numerical derivatives, we conduct synthetic data experiments of FWI to recover all the true viscoelastic parameters of targets. The results of the experiments support the validation of the derived Fréchet derivatives used for the 2-D/2.5-D seismic FWI in VEAM.

## METHODOLOGY

**2.5-D frequency-domain Fréchet derivatives in VEAM**

Methodologically speaking, the 2.5-D frequency-domain derivatives of the displacement tensor in VEAM may be obtained by extension of the elastic anisotropic case to its viscoelastic anisotropic counterpart. Note that the medium we are considering is anisotropic in both velocity (or moduli) and Q-factor. Here, we show the extension based on the 2.5-D frequency-domain derivatives of the displacement tensor in an elastic anisotropic medium below (Zhou and Greenhalgh, 2011):

$$\frac{\partial u_{\hat{s}\hat{g}}}{\partial m^{(e)}} = -f(\omega)F_y^{-1}\{\int_{\Omega_e}[\frac{\partial \tilde{c}_{ijkl}^{(e)}}{\partial m^{(e)}}\frac{\partial \bar{G}_{\hat{s}k}}{\partial x_l}\frac{\partial \bar{G}_{\hat{g}k}}{\partial x_i} + (\frac{\partial \tilde{c}_{2jk2}^{(e)}}{\partial m^{(e)}}k_y^2 - \frac{\partial \rho}{\partial m^{(e)}}\omega^2\delta_{jk}) + \\ + ik_y(\frac{\partial \tilde{c}_{ijk2}^{(e)}}{\partial m^{(e)}}\bar{G}_{\hat{s}k}\frac{\partial \bar{G}_{\hat{g}j}}{\partial x_i} - \frac{\partial \tilde{c}_{2jkl}^{(e)}}{\partial m^{(e)}}\bar{G}_{\hat{g}j}\frac{\partial \bar{G}_{\hat{s}k}}{\partial x_j})]d\Omega\} \tag{1}$$

where $u_{\hat{s}\hat{g}}$ is the component of the second-order displacement tensor. The subscripts '$\hat{s}$' and '$\hat{g}$' stand for the component directions of a source and a geophone, i.e., $\hat{s}, \hat{g}$=1, 2, 3 for the *x*-, *y*- and *z*-directions, respectively. The function $f(\omega)$ is the spectrum of the source wavelet. The vectors $\mathbf{\bar{G}}_{\hat{s}} = (\bar{G}_{\hat{s}1}, \bar{G}_{\hat{s}2}, \bar{G}_{\hat{s}3})$ and $\mathbf{\bar{G}}_{\hat{g}} = (\bar{G}_{\hat{g}1}, \bar{G}_{\hat{g}2}, \bar{G}_{\hat{g}3})$ are the wavenumber-domain Green's function tensors generated by the unit point-source vectors located at position coordinates $(x_{\hat{s}}, z_{\hat{s}})$ and $(x_{\hat{g}}, z_{\hat{g}})$, respectively. The scalar $m^{(e)} \in \{\rho^{(e)}(\mathbf{x}), \tilde{c}_{ijkl}^{(e)}(\mathbf{x}), \mathbf{x} \in \Omega_e\}$ is an independent model parameter that defines the physical property of a small subsurface block $\Omega_e$. Figure 1 is a sketch showing these vectors, tensors and model parameters. The symbol $F_y^{-1}$ denotes the inverse Fourier transform with respect to the wavenumber $k_y$, and the summation convention of the repeated subscripts {*ijkl*} is implied.

Applying the Gaussian quadrature grid method (Zhou et al, 2012) to the integral in equation 1, we have

$$\frac{\partial u_{\hat{s}\hat{g}}}{\partial m^{(e)}} = -f(\omega)w_p F_y^{-1}[\{\frac{\partial \tilde{c}_{ijkl}^{(e)}}{\partial m^{(e)}}\frac{\partial \bar{G}_{\hat{s}k}}{\partial x_l}\frac{\partial \bar{G}_{\hat{g}k}}{\partial x_i} + (\frac{\partial \tilde{c}_{2jk2}^{(e)}}{\partial m^{(e)}}k_y^2 - \frac{\partial \rho^{(e)}}{\partial m^{(e)}}\omega^2\delta_{jk} + \\ + ik_y(\frac{\partial \tilde{c}_{ijk2}^{(e)}}{\partial m^{(e)}}\bar{G}_{\hat{s}k}\frac{\partial \bar{G}_{\hat{g}j}}{\partial x_i} - \frac{\partial \tilde{c}_{2jkl}^{(e)}}{\partial m^{(e)}}\bar{G}_{\hat{g}j}\frac{\partial \bar{G}_{\hat{s}k}}{\partial x_j})\}J^{(e)}]_{\mathbf{x}_p}, \quad (\mathbf{x}_p \in \Omega_e). \tag{2}$$

where $\mathbf{x}_p = (x'_p, z'_p)$ and $w_p$ are the Gaussian abscissae and coefficients, respectively. $J^{(e)} = \partial(x,z)/\partial(x',z')$ is the Jacobian factor of the coordinate transformation for a complex geometry of the block $\Omega_e$, whose size is determined by the minimum wavelength of the simulated wavefields, subject to the spatial resolution requirement of the subsurface targets. This model discretization for 2-D FWI is called the constant block parameterization. Alternatively, one may employ the so-called constant point parameterization, i.e., setting up the elastic moduli and density functions by $\tilde{c}^{(e)}_{ijkl}(\mathbf{x}) = \tilde{c}^{(e)}_{ijkl}\delta(\mathbf{x}-\mathbf{x}_e)$ and $\rho^{(e)}(\mathbf{x}) = \rho^{(e)}\delta(\mathbf{x}-\mathbf{x}_e)$ respectively at the central points {$\mathbf{x}_e$, $e=1,2,...,M$} of small blocks, so that we have the derivatives: $\partial\tilde{c}^{(e)}_{ijkl}(\mathbf{x})/\partial m^{(e)} = \partial\tilde{c}^{(e)}_{ijkl}/\partial m^{(e)}\delta(\mathbf{x}-\mathbf{x}_e)$ and $\partial\rho^{(e)}(\mathbf{x})/\partial m^{(e)} = \partial\rho^{(e)}/\partial m^{(e)}\delta(\mathbf{x}-\mathbf{x}_e)$ for equation 1. Therefore, we have the derivatives for the constant point parameterization:

$$\frac{\partial u_{\hat{s}\hat{g}}}{\partial m^{(e)}} = -f(\omega)F_y^{-1}[\frac{\partial \tilde{c}^{(e)}_{ijkl}}{\partial m^{(e)}}\frac{\partial \bar{G}_{\hat{s}k}}{\partial x_l}\frac{\partial \bar{G}_{\hat{g}k}}{\partial x_i} + (\frac{\partial \tilde{c}^{(e)}_{2jk2}}{\partial m^{(e)}}k_y^2 - \frac{\partial \rho^{(e)}}{\partial m^{(e)}}\omega^2\delta_{jk}) + \\ + ik_y(\frac{\partial \tilde{c}^{(e)}_{ijk2}}{\partial m^{(e)}}\bar{G}_{\hat{s}k}\frac{\partial \bar{G}_{\hat{g}j}}{\partial x_i} - \frac{\partial \tilde{c}^{(e)}_{2jkl}}{\partial m^{(e)}}\bar{G}_{\hat{g}j}\frac{\partial \bar{G}_{\hat{s}k}}{\partial x_j})], \quad (3)$$

which varies point-by-point and gives the sensitivity of $u_{\hat{s}\hat{g}}$ to the specified parameter $m^{(e)} \in \{\rho^{(e)}, \tilde{c}^{(e)}_{ijkl}\}$ in strongly heterogeneous media.

It should be emphasized that both the adjoint method and the perturbation method can derive the above equations, which has been proved by Zhou and Greenhalgh (2009, 2011). The calculation of the Fréchet derivatives presented here is efficient, as it only needs two times of the forward modeling for the Green's function vectors $\bar{\mathbf{G}}_{\hat{s}}$ and $\bar{\mathbf{G}}_{\hat{g}}$.

To make equations 2 and 3 valid for VEAM, we employ the following fourth-order complex-valued stiffness tensor (Vavryčuk, 2007):

$$\tilde{c}_{ijkl}^{(e)} = c_{ijkl}^{(e)}(1 - i/Q_{ijkl}^{(e)}), \tag{4}$$

where $i = \sqrt{-1}$, and $c_{ijkl}^{(e)}$ and $Q_{ijkl}^{(e)}$ are the elastic moduli and the quality factors (Q-factor), which define the elasticity and viscoelasticity of the block $\Omega_e$, respectively. The Q-factors may be frequency-dependent or frequency-independent, and are constructed by different attenuation mechanical models, such as the generalized Kelvin-Voigt model, the generalized Maxwell model, or the generalized Zener (standard linear solid) model (Carcione et al., 1988; Casula and Carcione, 1992). Equation 4 does not use the Einstein summation convention of the repeated subscripts, so that it gives:

$$\frac{\partial \tilde{c}_{ijkl}^{(e)}}{\partial m^{(e)}} = (1 - \frac{i}{Q_{ijkl}^{(e)}})\frac{\partial c_{ijkl}^{(e)}}{\partial m^{(e)}} - \frac{ic_{ijkl}^{(e)}}{[Q_{ijkl}^{(e)}]^2}\frac{\partial Q_{ijkl}^{(e)}}{\partial m^{(e)}}, \tag{5}$$

which we can employ for equation 2 or 3 to calculate the derivative with respect to the independent parameter $m^{(e)} \in \{\rho^{(e)}, c_{ijkl}^{(e)}, Q_{ijkl}^{(e)}\}$. Note that as $k_y = 0$ m$^{-1}$ in equations 2 and 3, the 2.5-D frequency-domain derivatives become the 2-D derivatives, corresponding to the special line-source wavefield case, so that equations 2 and 3 remain valid for both point-source and line-source wavefields in a 2-D VEAM. Equations 2, 3 and 5 indicate that the derivatives of the displacement tensor require the $k_y$-domain Green's function vectors $\bar{\mathbf{G}}_{\hat{s}} = (\bar{G}_{\hat{s}1}, \bar{G}_{\hat{s}2}, \bar{G}_{\hat{s}3})$ and $\bar{\mathbf{G}}_{\hat{g}} = (\bar{G}_{\hat{g}1}, \bar{G}_{\hat{g}2}, \bar{G}_{\hat{g}3})$, and their gradients $\partial \bar{G}_{\hat{s}k}/\partial x_l$ and $\partial \bar{G}_{\hat{g}j}/\partial x_i$ in a heterogeneous VEAM model defined by different parameters $\{\rho^{(e)}, c_{ijkl}^{(e)}, Q_{ijkl}^{(e)}\}$ for all blocks $\{\Omega_e\}$ or points $\{\mathbf{x}_e\}$. Fortunately, Yang et al. (2020) has demonstrated the 2.5-D frequency-domain seismic wave modeling approach in such a VEAM, so that we apply it for the Green's function vectors $\bar{\mathbf{G}}_{\hat{s}}$ and $\bar{\mathbf{G}}_{\hat{g}}$, and then calculate the gradients $\partial \bar{G}_{\hat{s}k}/\partial x_l$ and $\partial \bar{G}_{\hat{g}j}/\partial x_i$ by numerical differentiation.

For the most general VEAM, equation 4 gives 21 complex-valued stiffness moduli (Carcione, 2007), but in practical situations, we only need to consider far fewer parameters than this. The viscoelastic isotropic and TTI type models are often encountered due to the symmetry of rock texture (Thomsen, 1986). The former is defined by two Lamé constants ($\tilde{\lambda}^{(e)}, \tilde{\mu}^{(e)}$) or the P- and S-wave velocities ($\tilde{\alpha}^{(e)}, \tilde{\beta}^{(e)}$). The latter is given by five independent moduli $\{\tilde{c}_{11}^{(e)}, \tilde{c}_{13}^{(e)}, \tilde{c}_{33}^{(e)}, \tilde{c}_{44}^{(e)}, \tilde{c}_{66}^{(e)}\}$ and the dip or tilt angle $\theta^{(e)}$ of the symmetry axis of the rock texture (see Figure 1) of each block. These two or six independent parameters are subsets of the 21 moduli. Considering the density $\rho^{(e)}$ and tilt angle $\theta^{(e)}$ to be real values (as opposed to complex) in VEAM, the derivatives $\partial u_{\hat{s}\hat{g}}/\partial \rho^{(e)}$ and $\partial u_{\hat{s}\hat{g}}/\partial \theta^{(e)}$ share the same form as in perfectly elastic media. Therefore, in the following sections we focus on the specified form of equation 5 for the stiffness moduli and Q-factors in the two common special cases.

**Two common cases**

1. Viscoelastic isotropic media

The viscoelastic isotropic moduli of a small block $\Omega_e$ are defined by (Aki and Richards, 1980)

$$c_{ijkl}^{(e)} = \tilde{\lambda}^{(e)} \delta_{ij}\delta_{kl} + \tilde{\mu}^{(e)}(\delta_{ik}\delta_{jl} + \delta_{jk}\delta_{il}) \tag{6}$$

where $\tilde{\lambda}^{(e)}$ and $\tilde{\mu}^{(e)}$ become the complex-valued Lamé parameters, such as

$$\tilde{\lambda}^{(e)} = \lambda^{(e)}(1 - \frac{i}{Q_\lambda^{(e)}}), \quad \tilde{\mu}^{(e)} = \mu^{(e)}(1 - \frac{i}{Q_\mu^{(e)}}), \tag{7}$$

where $\lambda^{(e)}$ and $\mu^{(e)}$ are the real-valued Lamé parameters to define the elasticity of the block $\Omega_e$. $Q_\lambda^{(e)}$ and $Q_\mu^{(e)}$ are the Q-factors to define the viscoelasticity of this block. Substituting equation 7 into 5 gives

$$\tilde{c}_{ijkl}^{(e)} = \lambda^{(e)}(1-\frac{i}{Q_\lambda^{(e)}})\delta_{ij}\delta_{kl} + \mu^{(e)}(1-\frac{i}{Q_\mu^{(e)}})(\delta_{ik}\delta_{jl} + \delta_{il}\delta_{jk}). \qquad (8)$$

Equation 8 leads to

$$\frac{\partial \tilde{c}_{ijkl}^{(e)}}{\partial m^{(e)}} = [(1-\frac{i}{Q_\lambda^{(e)}})\frac{\partial \lambda^{(e)}}{\partial m^{(e)}} - \frac{i\lambda^{(e)}}{[Q_\lambda^{(e)}]^2}\frac{\partial Q_\lambda^{(e)}}{\partial m^{(e)}}]\delta_{ij}\delta_{kl} + \\ [(1-\frac{i}{Q_\mu^{(e)}})\frac{\partial \mu^{(e)}}{\partial m^{(e)}} - \frac{i\mu^{(e)}}{[Q_\mu^{(e)}]^2}\frac{\partial Q_\mu^{(e)}}{\partial m^{(e)}}](\delta_{ik}\delta_{jl} + \delta_{il}\delta_{jk}) \qquad (9)$$

which shows four independent model parameters $m^{(e)} \in \{\lambda^{(e)}, \mu^{(e)}, Q_\lambda^{(e)}, Q_\mu^{(e)}\}$. Substituting equation 9 into 2 or 3, one obtains the derivatives for the viscoelastic isotropic case. Apparently, when the block becomes elastic, $Q_\lambda^{(e)} \to \infty$ and $Q_\mu^{(e)} \to \infty$, equation 9 gives

$$\frac{\partial \tilde{c}_{ijkl}^{(e)}}{\partial m^{(e)}} = \frac{\partial \lambda^{(e)}}{\partial m^{(e)}}\delta_{ij}\delta_{kl} + \frac{\partial \mu^{(e)}}{\partial m^{(e)}}(\delta_{ik}\delta_{jl} + \delta_{il}\delta_{jk}), \qquad (10)$$

which is the same as in the elastic isotropic case (Zhou and Greenhalgh, 2011), and equations 2 and 3 degenerate from VEAM to an elastic isotropic medium. Consequently, applying equation 9 for equation 2 or 3, one obtains the derivatives of the displacement tensor in arbitrary viscoelastic isotropic media. It is common to use the complex-valued P- and S-wave velocities, e.g. $\tilde{\alpha}^{(e)} = \alpha^{(e)}(1-\frac{i}{Q_\alpha^{(e)}})$ and $\tilde{\beta}^{(e)} = \beta^{(e)}(1-\frac{i}{Q_\beta^{(e)}})$ instead of equation 7. Here, $Q_\alpha^{(e)}$ and $Q_\beta^{(e)}$ are the Q-factors of the P- and S-wave. In this case, the five independent model parameters become $m^{(e)} \in \{\rho^{(e)}, \alpha^{(e)}, \beta^{(e)}, Q_\alpha^{(e)}, Q_\beta^{(e)}\}$ rather than $m^{(e)} \in \{\rho^{(e)}, \lambda^{(e)}, \mu^{(e)}, Q_\lambda^{(e)}, Q_\mu^{(e)}\}$. Appendix A gives the derivatives $\partial \lambda^{(e)}/\partial m^{(e)}$ and $\partial \mu^{(e)}/\partial m^{(e)}$, $m^{(e)} \in \{\rho^{(e)}, \alpha^{(e)}, \beta^{(e)}, Q_\alpha^{(e)}, Q_\beta^{(e)}\}$ for equation 9.

2. Viscoelastic TTI media

When the small block $\Omega_e$ in Figure 1 is a TTI solid, six independent model parameters $\{\theta^{(e)}, \tilde{c}_{1'1'}^{(e)}, \tilde{c}_{1'3'}^{(e)}, \tilde{c}_{3'3'}^{(e)}, \tilde{c}_{4'4'}^{(e)}, \tilde{c}_{6'6'}^{(e)}\}$ are required to define its viscoelastic anisotropic property. The angle $\theta^{(e)}$ gives the direction (see Fig. 1) of the symmetry axis of the rock texture. Therefore, equations 4 and 5 become

$$\tilde{c}_{p'q'}^{(e)} = c_{p'q'}^{(e)}(1 - i/Q_{p'q'}^{(e)}), \tag{11}$$

and

$$\frac{\partial \tilde{c}_{p'q'}^{(e)}}{\partial m^{(e)}} = (1 - i/Q_{p'q'}^{(e)})\frac{\partial c_{p'q'}^{(e)}}{\partial m^{(e)}} - \frac{ic_{p'q'}^{(e)}}{[Q_{p'q'}^{(e)}]^2}\frac{\partial Q_{p'q'}^{(e)}}{\partial m^{(e)}}, \tag{12}$$

where $c_{p'q'}^{(e)}$ and $Q_{p'q'}^{(e)}$ are the elastic moduli and Q-factors of block $\Omega_e$, and $(p'q') \in \{(1'1'), (1'3'), (3'3'), (4'4'), (6'6')\}$. Substituting equation 12 into 2 or 3, we obtain the derivatives $\partial u_{\hat{s}\hat{g}}/\partial m^{(e)}$, $m^{(e)} \in \{\theta^{(e)}, c_{1'1'}^{(e)}, c_{1'3'}^{(e)}, c_{3'3'}^{(e)}, c_{4'4'}^{(e)}, c_{6'6'}^{(e)}, Q_{1'1'}^{(e)}, Q_{1'3'}^{(e)}, Q_{3'3'}^{(e)}, Q_{4'4'}^{(e)}, Q_{6'6'}^{(e)}\}$. When $\theta^{(e)} = 0$, the 2-D viscoelastic TTI medium reduces to a 2-D viscoelastic vertically transversely isotropic (VTI) medium. As $Q_{p'q'}^{(e)} \to \infty$, $\forall(p'q')$, the block becomes a 2-D elastic TTI medium, and equation 12 becomes $\partial \tilde{c}_{p'q'}^{(e)}/\partial m^{(e)} = \partial c_{p'q'}^{(e)}/\partial m^{(e)}$. Accordingly, equations 12 and 2 or 3 are applicable for both elastic and viscoelastic TTI media.

## NUMERICAL EXAMPLES

To calculate the Fréchet derivatives given by equations 2 or 3, we carry out three steps:

(i) Discretize the subsurface domain $\Omega$ into either a set of non-overlapping small blocks $\Omega_e$, i.e., $\Omega = \bigcup_{e=1}^{M}\Omega_e$ or a set of points $\{\mathbf{x}_e \in \Omega, e=1,2,\ldots,M\}$, and then employ the twelve model parameters $m^{(e)} \in \{\rho^{(e)}, \theta^{(e)}, c_{1'1'}^{(e)}, c_{1'3'}^{(e)}, c_{3'3'}^{(e)}, c_{4'4'}^{(e)}, c_{6'6'}^{(e)}, Q_{1'1'}^{(e)}, Q_{1'3'}^{(e)}, Q_{3'3'}^{(e)}, Q_{4'4'}^{(e)}, Q_{6'6'}^{(e)}\}$ to define the viscoelasticity of the block $\Omega_e$ or at the points $\mathbf{x}_e$;

(ii) Apply 2.5-D frequency-domain wavefield modeling (Yang et al., 2020) to calculate the $k_y$-domain Green's function vectors $\bar{\mathbf{G}}_{\hat{s}} = (\bar{G}_{\hat{s}1}, \bar{G}_{\hat{s}2}, \bar{G}_{\hat{s}3})$ and $\bar{\mathbf{G}}_{\hat{g}} = (\bar{G}_{\hat{g}1}, \bar{G}_{\hat{g}2}, \bar{G}_{\hat{g}3})$ in the entire space of the viscoelastic isotropic or viscoelastic TTI medium;

(iii) With the Green's function vectors $\bar{\mathbf{G}}_{\hat{s}}$ and $\bar{\mathbf{G}}_{\hat{g}}$, we calculate the derivatives of the displacement tensor in terms of equations 2 and 9 or 12 for all the blocks ($\Omega_e, e = 1, 2, \ldots M$).

Step (i) is computationally negligible, and so is step (iii) once we have the Green's function vectors $\bar{\mathbf{G}}_{\hat{s}}$ and $\bar{\mathbf{G}}_{\hat{g}}$. After finishing these computations, we display the values of the derivatives in terms at the locations of the blocks or points. The spatial distributions of the derivative values in the model domain are often called the 'sensitivity patterns' of the displacement tensor $u_{\hat{s}\hat{g}}$ to the individual parameter of the subsurface. These sensitivity patterns indicate the sensitive areas (having relatively large values) of a seismic datum to each parameter of the subsurface, and their variations with the surveying geometry (i.e., positions of sources and geophones). Therefore, by comparing the sensitivity patterns for different data and different parameters, one may obtain the information on which parameter is the most sensitive and which datum (parts of the complex spectrum – frequency, real part, imaginary part plus source and geophone locations) has the most sensitive areas that cover the geological targets. Thus, one can select the most sensitive parameters and the parts of the acquisition geometry providing the most favorable spatial coverage for high-resolution subsurface imaging. All these information are crucial for an effective seismic FWI, and represent a requirement of a nonlinear local-search minimization algorithm, such as the Gauss-Newton method or Conjugate Gradient method, which requires the gradient and Hessian matrix of the data-misfit function. For example, to minimize the generalized least-squares data-misfit function, the gradient **g** and Hessian matrix **H** are often calculated by (Greenhalgh et al., 2006)

$$\begin{aligned} \mathbf{g} &= -(\frac{\partial \mathbf{u}}{\partial \mathbf{m}})^T \mathbf{W}_d [\mathbf{u}^{(ob)} - \mathbf{u}^{(syn)}] + \alpha \mathbf{W}_m (\mathbf{m} - \mathbf{m}_0) \\ \mathbf{H} &\approx (\frac{\partial \mathbf{u}}{\partial \mathbf{m}})^T \mathbf{W}_d (\frac{\partial \mathbf{u}}{\partial \mathbf{m}}) + \alpha \mathbf{W}_m \end{aligned} \tag{13}$$

where $\mathbf{W}_d$ and $\mathbf{W}_m$ are the weighting matrices of the data and model parameters respectively. One can set up in terms of a prior information on the data and subsurface model. The scalar $\alpha$ is the Tikhonov regularization parameter and makes the optimization stable and convergent. The vectors $\mathbf{u}^{(ob)}$ and $\mathbf{u}^{(syn)}$ are the observed and synthetic data of the displacement vector, and the derivative matrix $\partial \mathbf{u}/\partial \mathbf{m}$ is the Jacobian matrix whose components are calculated by equation (2) or (3). Equation (13) shows that the analytic derivatives given by equation (2) or (3), and equations (9) and (12) may be directly employed to form the gradient and Hessian matrix for the Gauss-Newton or Conjugate Gradient method for seismic FWI.

In order to view these sensitivity patterns, a 200 m × 200 m model domain is divided into 40 × 40 blocks, with one source and one receiver located at (-75 m, 100 m) and (75 m, 100 m), respectively. Four different homogenous background models/types of modeling were applied – 2-D elastic VTI, 2-D viscoelastic VTI, 2-D viscoelastic TTI and 2.5-D viscoelastic TTI. The model parameters are given in Table 1. These four sets of modeling results were used for comparison of the numerical solutions of the Green's function tensors with the analytic solutions (Casula and Carcione, 1992), to ensure that the computed $\overline{\mathbf{G}}_{\hat{s}}$ and $\overline{\mathbf{G}}_{\hat{g}}$ wavefield solutions are sufficiently accurate for computing the Fréchet derivatives of the displacement tensor. We use $\theta^{(e)} = 0°$ and $\theta^{(e)} = 30°(\forall \Omega_e)$ for the viscoelastic VTI and the viscoelastic TTI medium, respectively. As an example, the computational frequency of the wavefields is 50 Hz by applying a unit point-source spectrum $f(\omega) = 1$ ($\delta(t)$-pulse source). Other frequencies are also applicable for the wavefield computation (Yang et al. 2020).

The analytic and numerical solutions of the VTI and TTI models show only five non-zero components ($u_{11}, u_{13}, u_{22}, u_{31}, u_{33}$) in the central plane ($y$ = 0 m) (Vavryčuk, 2007; Yang et al., 2020; Zhao et al., 2017). Therefore, the derivatives $\partial u_{\hat{s}\hat{g}}/\partial m^{(e)}$ also have five non-zero components. To display the sensitivity patterns, we applied a homogenous background, i.e., $\forall \Omega_e$, $m^{(e)} = m \in \{\rho, \theta, c_{1'1'}, c_{1'3'}, c_{3'3'}, c_{4'4'}, c_{6'6'}, Q_{1'1'}, Q_{1'3'}, Q_{3'3'}, Q_{4'4'}, Q_{6'6'}\}$ which means the same twelve model parameters are used in each block. Due to the homogenous background, we found that the derivatives $\partial u_{13}/\partial m$ and $\partial u_{31}/\partial m$ have the same sensitive patterns and the real parts $\text{Re}[\partial u_{sg}/\partial m]$ and the imaginary parts $\text{Im}[\partial u_{sg}/\partial m]$ are very similar. Accordingly, we only show the real parts of the four derivatives: $\text{Re}[\partial u_{11}/\partial m]$, $\text{Re}[\partial u_{13}/\partial m]$, $\text{Re}[\partial u_{22}/\partial m]$ and Re[$\partial u_{33}/\partial m$] for the twelve independent parameters. In order to distinguish the derivatives in the 2-D viscoelastic VTI and TTI media from the 2-D elastic VTI and TTI media, and the 2.5-D derivatives from the 2-D derivatives, we show all the results for the four different models given in Table 1.

Figure 2 gives the numerical results of the real parts of the density derivatives $\partial u_{\hat{s}\hat{g}}/\partial \rho$ for the four non-zero components ($u_{11}, u_{13}, u_{22}, u_{33}$) in the four models. There are 16 panels in the figure, with the 4 columns corresponding to the 4 different models/types of modeling and the 4 rows corresponding to the 4 different derivatives. All panels have a quasi-elliptical shape, with foci coinciding with the source and receiver locations. The direct zone between source and receiver is called the first Fresnel zone. The intricate phases around the first Fresnel zone are observed due to the coupling between different wave modes {qP, qSV} except for the panels of $\partial u_{22}/\partial \rho$, which exhibit the first Fresnel zone and superposition (constructive/ destructive interference) between the source (forward) and receiver (adjoint) wavefields. Comparing the 1st column (2-D elastic VTI) with the 2nd column (2-D viscoelastic VTI), one can see that the magnitude of the derivatives in the viscoelastic case is a somewhat weaker than that in the elastic case. This is due to the attenuation of the wave energy. Comparing the 2nd column (2-D viscoelastic VTI, θ =0°) with the 3rd column (2-D viscoelastic TTI, $\theta$=30° ), one finds significant phase changes in $\partial u_{22}/\partial \rho$, whereas the other components in the TTI case show a slightly wider first Fresnel zone than that in the VTI medium. This is due to a 30° rotation of the symmetry axis. Comparing the 3rd column (2-D viscoelastic TTI) with the 4th column (2.5-D viscoelastic TTI), one can observe a distinctly weaker amplitude of the point-source wavefield than that of the line-source wavefield. The former is smaller by two orders of magnitude compared to all 2-D cases (1st – 3rd columns).

Figure 3 illustrates the real parts of the tilt angle derivatives $\partial u_{\hat{s}\hat{g}}/\partial\theta$ for the four non-zero components ($u_{11}, u_{13}, u_{22}, u_{33}$) in the four models/types of modeling. The 1st column (2-D elastic VTI) shows that all derivatives essentially vanish except for $\partial u_{22}/\partial\theta$, whereas the 2nd column (2-D viscoelastic VTI) indicates that $\partial u_{11}/\partial\theta$, $\partial u_{13}/\partial\theta$ and $\partial u_{33}/\partial\theta$ are non-zero but exhibit very weak magnitudes because of the effects of the Q-factors ($Q_{11}$=50, $Q_{13}$=35, $Q_{33}$=50, $Q_{44}$=40). Another feature shown in Figure 3 is that the derivatives of $\partial u_{22}/\partial\theta$ exhibit rotational symmetries in the two TTI cases. The last two columns in Figure 3 indicate similar sensitivity patterns but the amplitudes of the point-source derivatives are less by at least two orders of magnitude than the line-source results, similar to what we observed in the last two columns of Figure 2.

Figures 4 to 8 show the real parts of the moduli derivatives $\partial u_{\hat{s}\hat{g}}/\partial m$ for the four non-zero components ($u_{11}, u_{13}, u_{22}, u_{33}$) and the five independent elastic moduli $m \in \{c_{1'1'}, c_{1'3'}, c_{3'3'}, c_{4'4'}, c_{6'6'}\}$ in the four models. Comparing them with Figures 2-3, we can see that the patterns and magnitudes of all five Fréchet derivatives with respect to the elastic moduli are different from the Fréchet derivatives with respect to density $\rho$ and the derivative with respect to the dip angle $\theta$. The magnitudes of the derivatives $\partial u_{\hat{s}\hat{g}}/\partial c_{p'q'}$ ($\approx 10^{-21} \sim 10^{-24}$) are much smaller than $\partial u_{\hat{s}\hat{g}}/\partial \rho$ and $\partial u_{\hat{s}\hat{g}}/\partial \theta$ ($\approx 10^{-12} \sim 10^{-17}$) shown in Figures 2 and 3. This is because the elastic moduli given in Table 1 are 1.00 ~ 6.26 ×$10^9$ Pa, from which one can see that the 1st-order change of the displacement tensor $\Delta u_{\hat{s}\hat{g}} = (\partial u_{\hat{s}\hat{g}}/\partial c_{p'q'})\Delta c_{p'q'}$ is about $10^{-12} \sim 10^{-15}$, which can be contrasted with $\Delta u_{\hat{s}\hat{g}} = (\partial u_{\hat{s}\hat{g}}/\partial \rho)\Delta \rho$ or $\Delta u_{\hat{s}\hat{g}} = (\partial u_{\hat{s}\hat{g}}/\partial \theta)\Delta \theta$ ($\approx 10^{-12} \sim 10^{-17}$). Therefore, the variation magnitudes of the displacement-tensor $\Delta u_{\hat{s}\hat{g}}$ caused by perturbations of the moduli $\Delta c_{p'q'}$, density $\Delta \rho$ and dip angle $\Delta \theta$ are comparable, but their sensitivity patterns are different. The different sensitivity patterns shown in Figures 4 to 8 indicate that the seven model parameters have different effects on the displacement tensor, e.g. one can observe that because the component $u_{22}$ only depends on $c_{4'4'}$ and $c_{6'6'}$ in all 2-D cases (the 1st to 3rd columns), the derivatives $\partial u_{22}/\partial c_{4'4'}$ and $\partial u_{22}/\partial c_{6'6'}$ are non-zero whereas the others vanish. The magnitudes of the derivatives in the viscoelastic cases (the 2nd-3rd columns) are weaker than that in elastic case (the 1st column), and the point-source results (the 4th column) are the weakest of all four models. However, in the 2.5-D case (see the 4th column in Figures 4 to 8), none of the derivatives vanish. This indicates the significant difference between a line-source and a point-source. Looking at Figures 4 and 5, one finds that the derivatives with respect to $c_{1'1'}$ and $c_{1'3'}$ show very similar patterns, but the other derivatives with respect to $c_{3'3'}$, $c_{4'4'}$, and $c_{6'6'}$ exhibit quite different patterns.

Due to the non-existence of the Q-factors in perfectly elastic media, we only have to consider the Q derivatives $\partial u_{\hat{s}\hat{g}}/\partial c_{p'q'}$ in the three viscoelastic models given in Table

1. Figures 9 to 13 give the real parts of the derivatives $\partial u_{\hat{s}\hat{g}}/\partial Q_{p'q'}$ for the four non-zero components $(u_{11}, u_{13}, u_{22}, u_{33})$ and the five independent Q-factors $(Q_{1'1'}, Q_{1'3'}, Q_{3'3'}, Q_{4'4'}, Q_{6'6'})$, Applying the derivative chain law to $u_{\hat{s}\hat{g}}(\tilde{c}_{p'q'})$ and equation 11, we have

$$\begin{aligned}\frac{\partial u_{\hat{s}\hat{g}}}{\partial Q_{p'q'}} &= \frac{ic_{p'q'}}{[Q_{p'q'}]^2}\frac{\partial u_{\hat{s}\hat{g}}}{\partial \tilde{c}_{p'q'}} \\ &= \frac{ic_{p'q'}}{[Q_{p'q'}]^2(1-iQ_{p'q'}^{-1})}\frac{\partial u_{\hat{s}\hat{g}}}{\partial c_{p'q'}} \\ &= \frac{-c_{p'q'}(Q_{p'q'}^{-1}-i)}{(1+Q_{p'q'}^2)}\frac{\partial u_{\hat{s}\hat{g}}}{\partial c_{p'q'}}\end{aligned} \quad (14)$$

which implies

$$\begin{aligned}\text{Re}[\frac{\partial u_{\hat{s}\hat{g}}}{\partial Q_{p'q'}}] &= \frac{-c_{p'q'}}{Q_{p'q'}(1+Q_{p'q'}^2)}\text{Re}[\frac{\partial u_{\hat{s}\hat{g}}}{\partial c_{p'q'}}] - \frac{c_{p'q'}}{(1+Q_{p'q'}^2)}\text{Im}[\frac{\partial u_{\hat{s}\hat{g}}}{\partial c_{p'q'}}], \\ \text{Im}[\frac{\partial u_{\hat{s}\hat{g}}}{\partial Q_{p'q'}}] &= \frac{c_{p'q'}}{(1+Q_{p'q'}^2)}\text{Re}[\frac{\partial u_{\hat{s}\hat{g}}}{\partial c_{p'q'}}] - \frac{c_{p'q'}}{Q_{p'q'}(1+Q_{p'q'}^2)}\text{Im}[\frac{\partial u_{\hat{s}\hat{g}}}{\partial c_{p'q'}}].\end{aligned} \quad (15)$$

This indicates that the real and imaginary parts of $\partial u_{\hat{s}\hat{g}}/\partial Q_{p'q'}$ are given by combinations of the real and imaginary parts of $\partial u_{\hat{s}\hat{g}}/\partial c_{p'q'}$. Therefore, $\partial u_{\hat{s}\hat{g}}/\partial Q_{p'q'}$ should have similar sensitivity patterns to $\partial u_{\hat{s}\hat{g}}/\partial c_{p'q'}$. Comparing Figures 9 to 13 with the 2$^{nd}$ ~ 4$^{th}$ columns in Figures 4 to 8, one can see that all the sensitivity patterns of the two sets are indeed similar except for their magnitudes. The magnitude differences of the derivatives can be explained by equation 12. In practical applications of FWI, mitigating the influence of the magnitude differences of the derivatives (sensitivities) is often necessary. One may apply scaled gradients or normalized derivatives to the Newton or conjugate gradient optimization algorithm in 2-D FWI inversion, so that these independent parameters can be effectively determined by the iterative optimization process.

To partially verify the capability of the derivatives given in equations 2 and 12, we apply them to 2-D frequency-domain FWI that individually inverts twelve model parameters of an anomalous block target embedded at the center of a homogeneous viscoelastic TTI background. The whole model size is 200 m × 200 m, and the target size is 30 m × 30 m. The model parameters are 10 percent higher or lower than the background values. The observed data are generated by 10 sources on the top of the model and recorded by 40 receivers arranged at uniform spacing along the four edges which surround the model. A Ricker wavelet with 15 Hz center frequency is applied. Such a simple model and the ideal 4-sided surveying geometry are employed just for validation of the derivatives in viscoelastic TTI media. We started the 2-D FWI with the homogeneous TTI medium and reconstructed the perturbed model parameters one by one with the synthetic data. Figure 14 shows the central-line plots of the inversion results (indicated by dotted lines) of the twelve model parameters. The stable $l$-BFGS inversion method is used. The frequencies of 7, 12, 25, 37, 50 Hz and 10 iterations per frequency are used to accomplish these inversions. From these plots, one can observe that all the twelve parameters are re-constructed correctly and accurately with this ideal recording geometry. The plots of five quality factors ($Q_{1'1'}, Q_{1'3'}, Q_{3'3'}, Q_{4'4'}, Q_{6'6'}$) are very similar to that of the corresponding real moduli ($c_{1'1'}, c_{1'3'}, c_{3'3'}, c_{4'4'}, c_{6'6'}$), which can be explained by the linear relationships of their derivatives (equation 12). All these results validate the derived derivatives of the displacement tensor for viscoelastic TTI media.

Another multi-target model is inverted to present impacts from four common seismic surveying geometries — surface, crosshole, vertical seismic profiling (VSP), and VSP+surface surveys. For these experiments, we employ 10 sources on the surface and 10 receivers at equal spacing along each edges of the model according to these surveying geometries. The model includes two square anomalous targets having different $Q_{1'1'}$ values (see Figure 15a) from the background 2-D viscoelastic TTI medium given in Table 1. Similarly, five frequencies (7, 12, 25, 37, 50 Hz) are used and 10 iterations are implemented for each frequency within the *l*-BFGS inversion method. Figures 15b-e show the inverted results under the four seismic surveys. From these results one can easily recognize the anomalous $Q_{1'1'}$ values of the two targets. Obviously, more extensive observation angle gets more accurate result of the targets. These different surveys can also be seen as with different offset lengths. The surface survey has the shortest offset, so it has the worst result shown in Figure 15b, vice versa the VSP+surface survey (Figure 15e) yields the best images in these four results. From Figures 15b and 15c, one can also see the artificial errors present 30° inclination from the *z*-axis which is consistent with the rotation angle of the background TTI medium, while Figures 15d and 15e show significantly less artificial errors due to extension of the surveying offset in the vertical and horizontal directions. Similarly, the $Q_{1'1'}$ inversions can be carried out to any other Q-factor. Such multi-target Q-factor inversion once again verifies the Fréchet derivatives given by equations (3) and (12) applicable for the FWI to obtain Q-factor image of a 2-D viscoelastic TTI media.

## DISCUSSION

In practice, we may not have such a simple model and a prior information about all these model parameters, nor an ideal surveying geometry in the first inversion experiment, but it is worthy of demonstrating the effectiveness of the derived Fréchet derivatives to recover each of the viscoelastic anisotropic parameters. The multi-target Q-factor inversions with four common seismic surveys simulated possible applications in practice and show the capability of using the analytic Fréchet derivatives in seismic FWI. It clearly indicates that the FWI needs long-offset observed data to reduce the artificial errors in the inverted images due to the incompleteness of the data observation. We have analyzed the Fréchet derivatives involving the ground displacement vector $u$ as the observable. In practice, it is far more common to record particle velocity vector $v$ with a geophone or pressure with a hydrophone $p$. It is a simple matter to convert from one to the other using, for example, the frequency-domain relationship $v = i\omega u$ and the differential chain rule $\frac{\partial v}{\partial m} = \frac{\partial u}{\partial m} \cdot \frac{\partial v}{\partial u}$. Conducting Multi-parameter seismic FWI with these common survey configurations is the next step of this research.

The possible strategies for the practical applications may involve four key aspects: (1) focus on the most sensitive model parameters of the common seismic surveys; (2) convert the point-source wavefield data into the line-source data; (3) employ multi-components of the full-waveform data, and (4) apply geological structure constraints, such as zero cross-gradients (Aki and Richards, 1980) and the maximum cross-correlation (Greenhalgh et al., 2006) to the inversion.

**CONCLUSIONS**

This paper has presented the frequency-domain Fréchet derivatives of the displacement tensor for 2-D/2.5-D seismic FWI in arbitrary viscoelastic anisotropic media (VEAM). Two common cases (a viscoelastic isotropic medium and a viscoelastic TTI medium) are specifically demonstrated. The formulations indicate that the derivatives require the two $k_y$-domain Green's function tensors $\overline{\mathbf{G}}_{\hat{s}}$ and $\overline{\mathbf{G}}_{\hat{g}}$, which are the wavefields corresponding to a unit-vector line-source or point-source at the source and geophone locations, respectively. Both can be obtained by a 2.5-D frequency-domain wave modeling method, so that the Fréchet derivatives of the displacement tensor can be efficiently calculated in an arbitrary VEAM.

Our numerical examples give the comparison of the sensitivity patterns of the displacement tensor in four homogenous models: 2-D elastic and viscoelastic VTI models and 2-D and 2.5-D viscoelastic TTI models. From these results, one may observe four characteristics: (1) the sensitivity patterns in the 2-D viscoelastic VTI medium are almost the same as in the 2-D elastic VTI medium, but the former has weaker magnitudes than the latter due to the viscoelastic damping of the medium, (2) the dip or tilt angle in the TTI medium makes the sensitivity patterns significantly distorted from the results of the VTI medium, (3) the sensitivities of the 2.5-D displacement tensor are weaker than those of the 2-D case by at least two orders of magnitude, due to the salient difference between the line- and point-source, and (4) the derivatives of the displacement tensor with respect to the Q-factors are calculated by linear combinations of the real and imaginary parts of the derivatives with respect to the elastic moduli, and their sensitivity patterns are very similar to each other, but their magnitudes have significant differences.

Our inversion examples show that applying the Fréchet derivatives of the displacement tensor with an ideal survey geometry to the frequency-domain FWI enables us to individually recover all twelve model parameters of a square block target embedded in a homogeneous viscoelastic medium. These results validate the applicability of the numerical computations of the Fréchet derivatives of the displacement tensor for viscoelastic TTI media, and the numerical derivatives may be employed for a local-search optimization algorithm, such as the Gauss-Newton method and the conjugate gradient method for 2-D or 2.5-D FWI. Our future research will be devoted to multi-parameter and multi-component seismic FWI for a complex viscoelastic VTI or TTI medium using the common seismic survey geometries.

**Appendix A. Inter-relationships between moduli, velocity, density and Q parameters and their mutual derivatives for isotropic elastic media**

According to the P- and S-wave velocities for an elastic medium, i.e., $\alpha^2 = (\lambda + 2\mu)/\rho$ and $\beta^2 = \mu/\rho$ and their extension to a viscoelastic medium: $\alpha^2 \Rightarrow \alpha^2(1 - iQ_\alpha^{-2})$, $\beta^2 \Rightarrow \beta^2(1 - iQ_\beta^{-2})$, $\lambda \Rightarrow \lambda(1 - iQ_\lambda^{-1})$ and $\mu \Rightarrow \mu(1 - iQ_\mu^{-1})$, one may find the following relationship between the parameterizations $\{\lambda, \mu, Q_\lambda, Q_\mu\}$ and

$\{\alpha, \beta, Q_\alpha, Q_\beta\}$:

$$\lambda = \rho[\alpha^2(1-Q_\alpha^{-2}) - 2\beta^2(1-Q_\beta^{-2})],$$
$$\mu = \rho\beta^2(1-Q_\beta^{-2}),$$
$$Q_\lambda = \frac{\alpha^2(1-Q_\alpha^{-2}) - 2\beta^2(1-Q_\beta^{-2})}{2\alpha^2 Q_\alpha^{-1} - 4\beta^2 Q_\beta^{-1}}, \quad (A1)$$
$$Q_\mu = \frac{1-Q_\beta^{-2}}{2Q_\beta^{-1}}.$$

Equation A1 shows that the parameters $\{\lambda, \mu, Q_\lambda, Q_\mu\}$ may be calculated from the common parameters $\{\rho, \alpha, \beta, Q_\alpha, Q_\beta\}$, from which we have the following derivatives:

$$\frac{\partial \lambda}{\partial \rho} = \alpha^2(1-Q_\alpha^{-2}) - 2\alpha^2(1-Q_\beta^{-2}), \quad \frac{\partial \lambda}{\partial \alpha} = 2\alpha\beta(1-Q_\alpha^{-2}),$$
$$\frac{\partial \lambda}{\partial \beta} = -4\rho\beta(1-Q_\beta^{-2}), \quad \frac{\partial \lambda}{\partial Q_\alpha} = 2\rho\alpha^2 Q_\alpha^{-3},$$
$$\frac{\partial \lambda}{\partial Q_\beta} = -4\rho\beta^2 Q_\beta^{-3}, \quad \frac{\partial \mu}{\partial \rho} = \beta^2(1-Q_\beta^{-2}), \quad (A2)$$
$$\frac{\partial \mu}{\partial \beta} = 2\rho\beta(1-Q_\beta^{-2}), \quad \frac{\partial \mu}{\partial Q_\beta} = 2\rho\beta^2 Q_\beta^{-3}.$$

$$\frac{\partial Q_\lambda}{\partial \alpha} = \frac{2\alpha\beta^2(Q_\alpha^{-1} - Q_\beta^{-1})(1 + Q_\alpha^{-1} Q_\beta^{-1})}{(\alpha^2 Q_\alpha^{-1} - 2\beta^2 Q_\beta^{-1})},$$
$$\frac{\partial Q_\lambda}{\partial \beta} = \frac{2\alpha\beta^2(Q_\beta^{-1} - Q_\alpha^{-1})(1 + Q_\alpha^{-1} Q_\beta^{-1})}{(\alpha^2 Q_\alpha^{-1} - 2\beta^2 Q_\beta^{-1})}$$
$$\frac{\partial Q_\lambda}{\partial Q_\alpha} = \alpha^2 Q_\alpha^{-2} \frac{\alpha^2(1+Q_\alpha^{-2}) - 2\beta^2(1-Q_\beta^{-2} + 2Q_\alpha^{-1} Q_\beta^{-1})}{2(\alpha^2 Q_\alpha^{-1} - 2\beta^2 Q_\beta^{-1})}, \quad (A3)$$
$$\frac{\partial Q_\lambda}{\partial Q_\beta} = -\beta^2 Q_\beta^{-2} \frac{\alpha^2(1+Q_\alpha^{-2} + 2Q_\alpha^{-1} Q_\beta^{-1}) - 2\beta^2(1+Q_\beta^{-2})}{2(\alpha^2 Q_\alpha^{-1} - 2\beta^2 Q_\beta^{-1})},$$
$$\frac{\partial Q_\mu}{\partial Q_\beta} = \frac{Q_\beta^{-2} + 1}{2}.$$

Substituting A2 and A3 into equations 9 and 2 or 3, one obtains the Fréchet derivatives $\{\partial u_{\hat{s}\hat{g}}/\partial m^{(e)}\}$, $m^{(e)} \in \{\rho^{(e)}, \alpha^{(e)}, \beta^{(e)}, Q_\alpha, Q_\beta\}$. Similarly, one may deduce the derivatives for the independent model parameters $m \in \{\rho, \kappa, \mu, Q_\kappa, Q_\mu\}$ in terms of their inter-relationships. Here, $\kappa$ and $Q_\kappa$ are the bulk modulus and its corresponding quality factor (Yang et al., 2016).

**Table 1.** Models and their model parameters

| Model | Density | Elastic moduli | Quality factors | Dipping angle |
|---|---|---|---|---|
| 2-D elastic VTI | | $c_{1'1'}$=6.25, | $Q_{p'q'}=\infty$ | $\theta^{(e)}=0°$ |
| 2-D viscoelastic VTI | | $c_{1'3'}$=4.25, | $Q_{1'1'}$=50, | $\theta^{(e)}=0°$ |
| 2-D viscoelastic TTI | $\rho$=1000 (kg/m³) | $c_{3'3'}$=6.25, | $Q_{1'3'}$=35, | $\theta^{(e)}=30°$ |
| 2.5-D viscoelastic TTI | | $c_{4'4'}$=1.00, $c_{6'6'}$=2.25. (GPa) | $Q_{3'3'}$=50, $Q_{4'4'}$=40, $Q_{6'6'}$=20. | $\theta^{(e)}=30°$ |

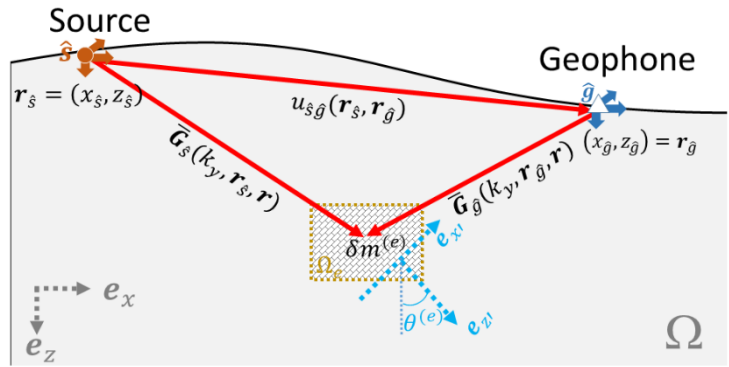

Figure 1. Sketch of a component $u_{\hat{s}\hat{g}}$ of the displacement tensor, the Green's function vectors $\bar{\mathbf{G}}_{\hat{s}} = (\bar{G}_{\hat{s}1}, \bar{G}_{\hat{s}2}, \bar{G}_{\hat{s}3})$ and $\bar{\mathbf{G}}_{\hat{g}} = (\bar{G}_{\hat{g}1}, \bar{G}_{\hat{g}2}, \bar{G}_{\hat{g}3})$ and the model parameter perturbation $\delta\mathbf{m}^{(e)} = \left(\delta m_1^{(e)}, \delta m_2^{(e)}, \dots, \delta m_\nu^{(e)}\right)$ of a block $\Omega_e$.

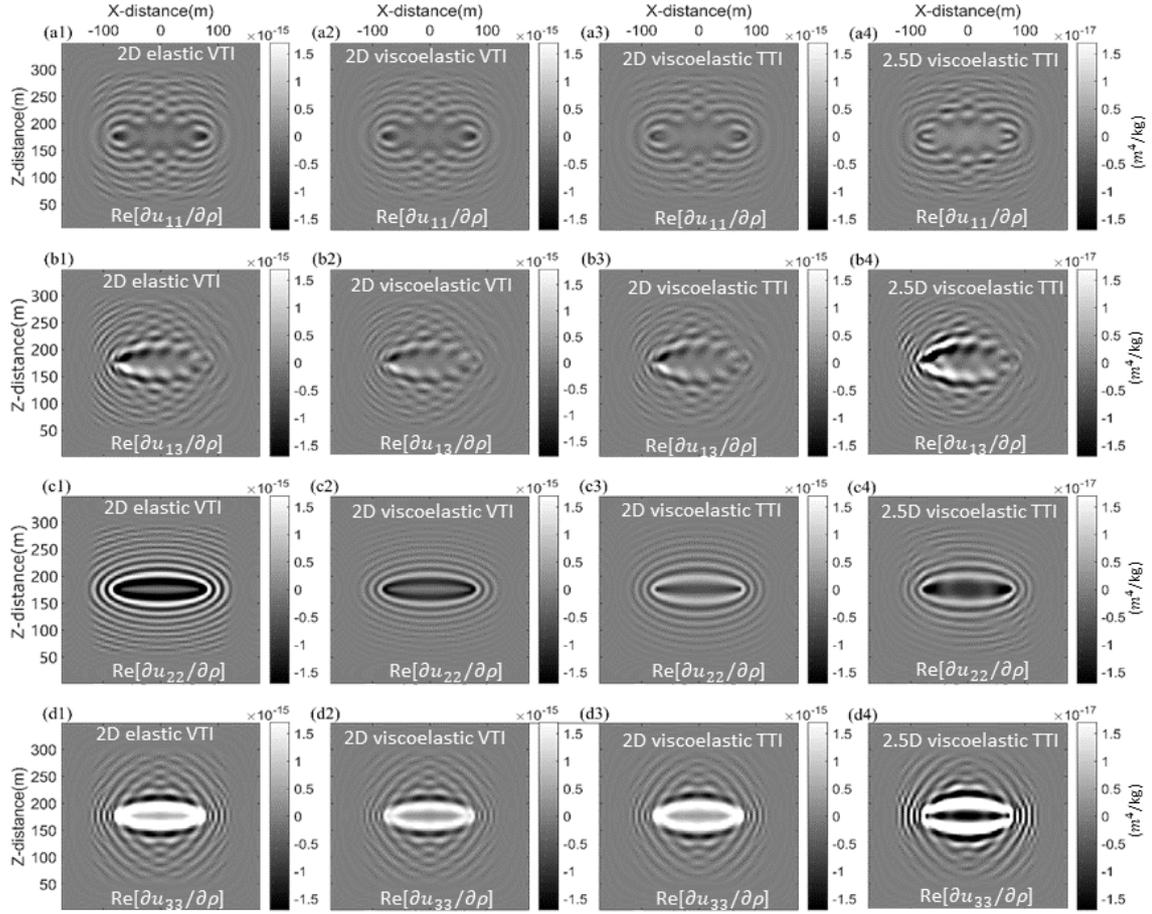

Figure 2. Numerical results of $\text{Re}[\partial u_{\hat{s}\hat{g}}/\partial \rho]$ in four homogeneous models: (1) 2-D elastic VTI, (2) 2-D viscoelastic VTI, (3) 2-D viscoelastic TTI ($\theta=30°$) and (4) 2.5-D viscoelastic TTI ($\theta=30°$).

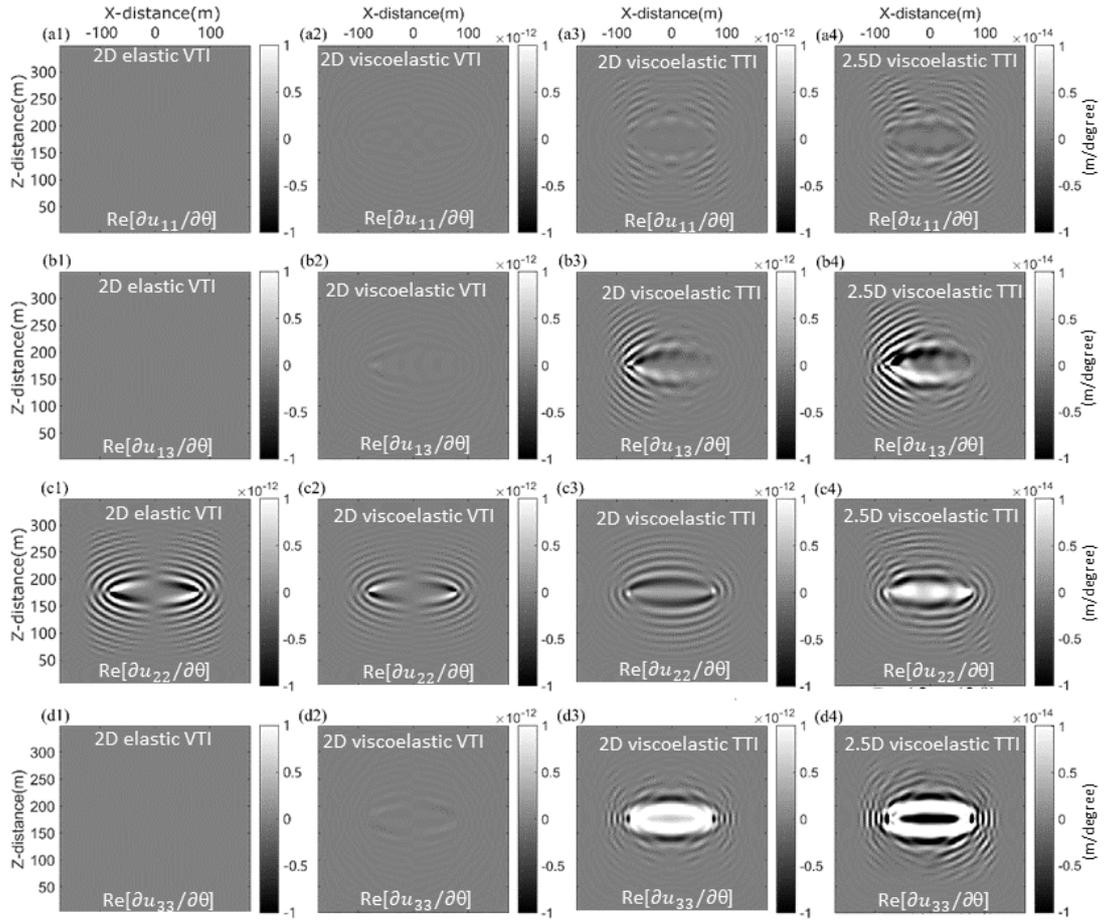

Figure 3. Numerical results of $\text{Re}[\partial u_{\hat{s}\hat{g}}/\partial \theta]$ in four homogeneous media: (1) 2-D elastic VTI, (2) 2-D viscoelastic VTI, (3) 2-D viscoelastic TTI ($\theta=30°$) and (4) 2.5-D viscoelastic TTI ($\theta=30°$).

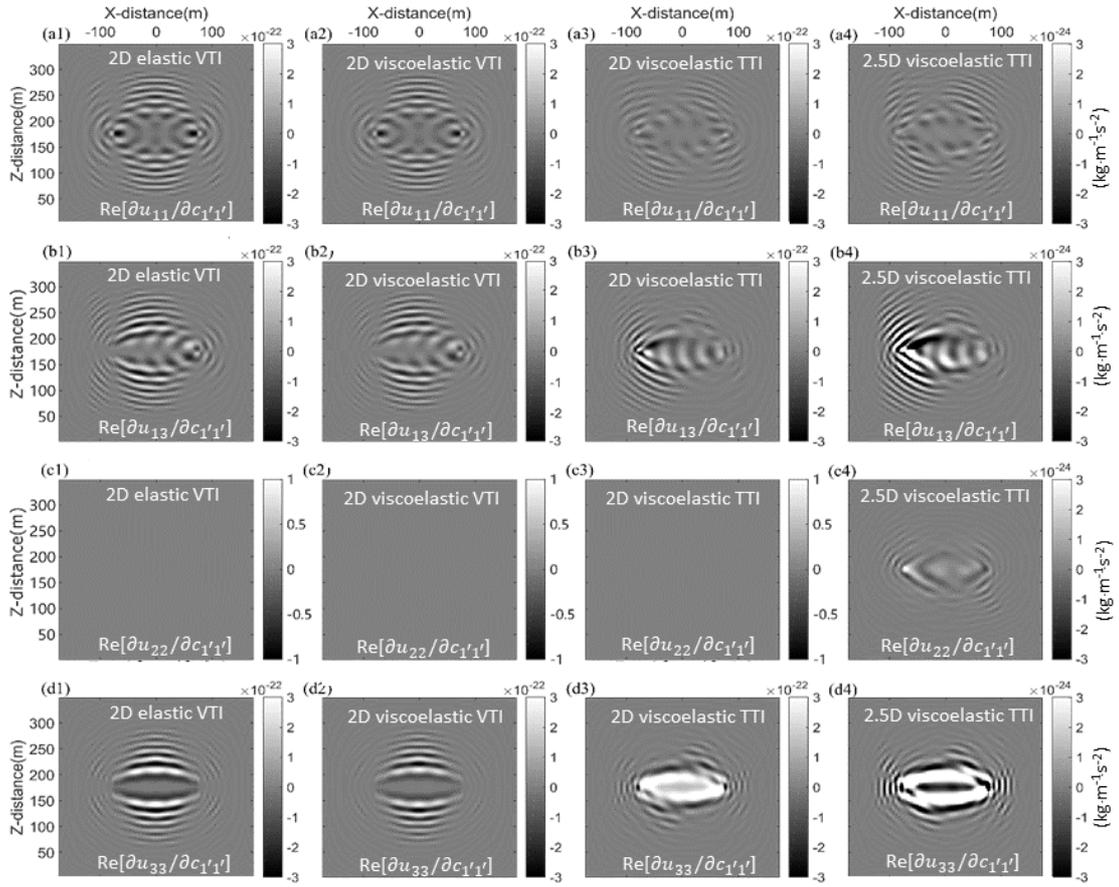

Figure 4. Numerical results of $\text{Re}[\partial u_{\hat{s}\hat{g}}/\partial c_{1'1'}]$ in four homogeneous media: (1) 2-D elastic VTI, (2) 2-D viscoelastic VTI, (3) 2-D viscoelastic TTI ($\theta=30°$) and (4) 2.5-D viscoelastic TTI ($\theta=30°$).

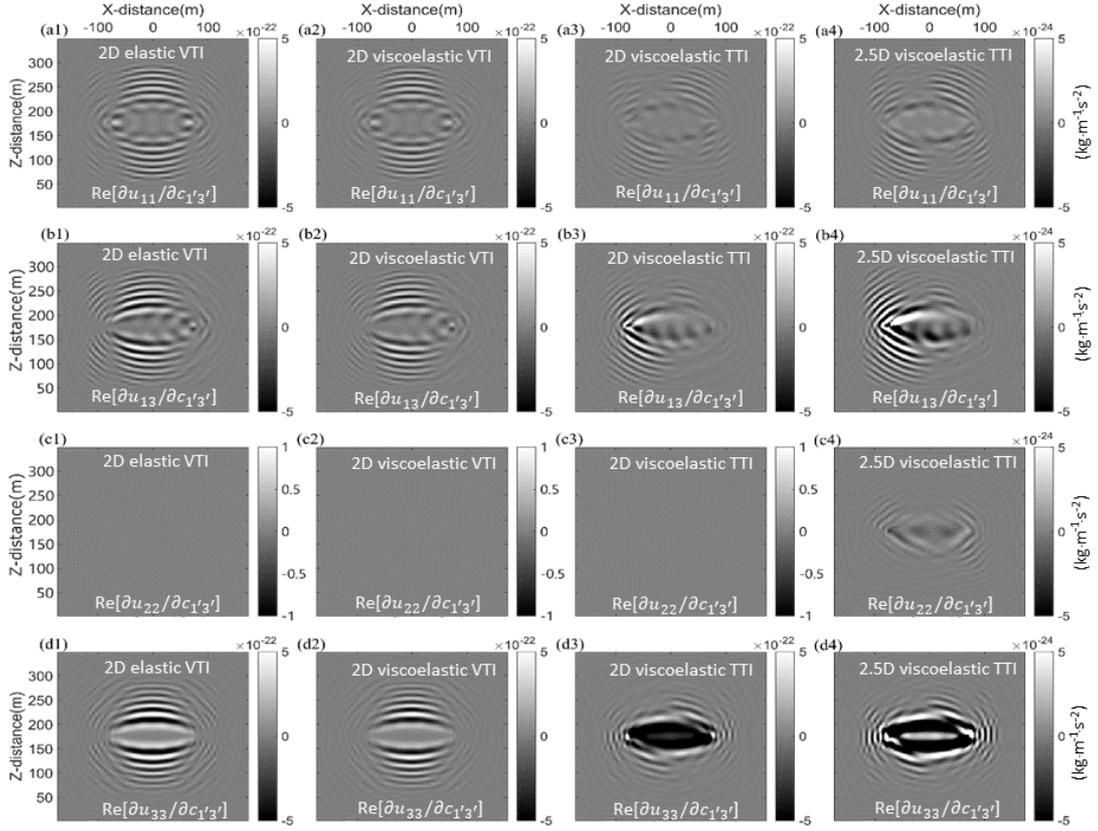

Figure 5. Numerical results of $\mathrm{Re}[\partial u_{\hat{s}\hat{g}}/\partial c_{1'3'}]$ in four homogeneous media: (1) 2-D elastic VTI, (2) 2-D viscoelastic VTI, (3) 2-D viscoelastic TTI ($\theta=30°$) and (4) 2.5-D viscoelastic TTI ($\theta=30°$).

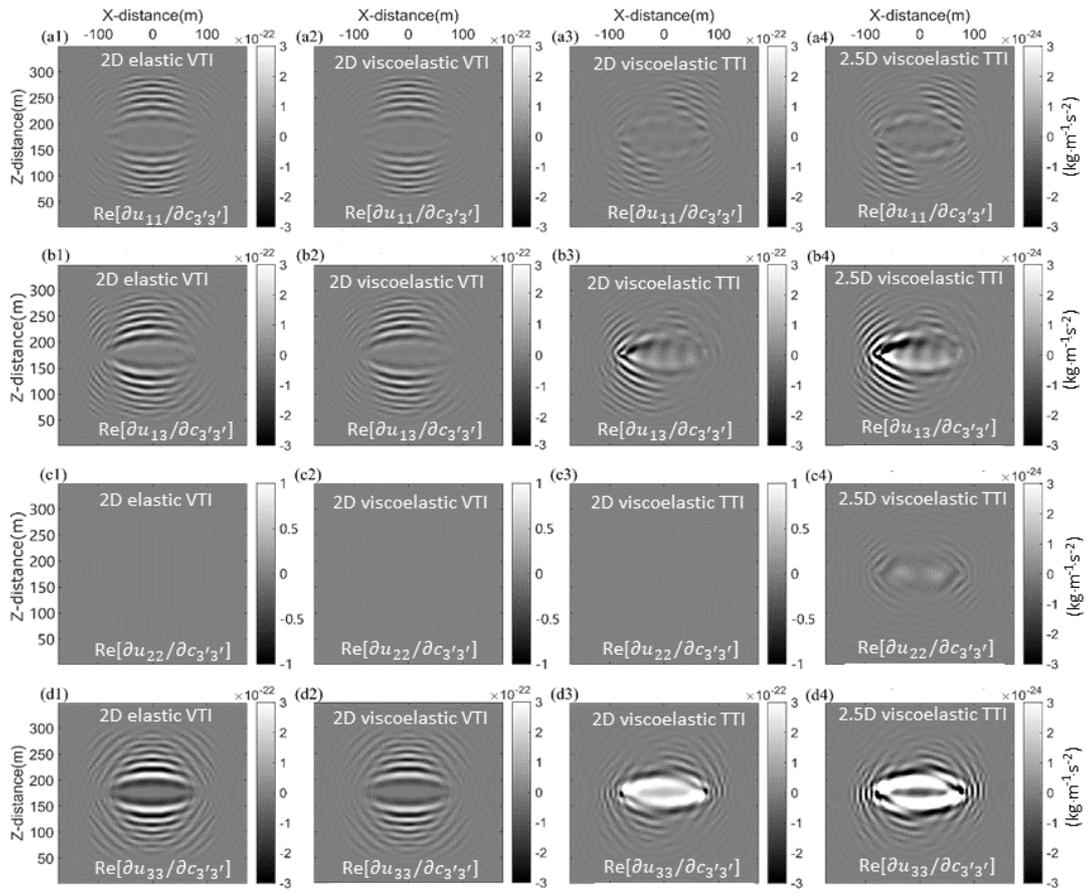

Figure 6. Numerical results of $\text{Re}[\partial u_{\hat{s}\hat{g}}/\partial c_{3'3'}]$ in four homogeneous media: (1) 2-D elastic VTI, (2) 2-D viscoelastic VTI, (3) 2-D viscoelastic TTI ($\theta=30°$) and (4) 2.5-D viscoelastic TTI ($\theta=30°$).

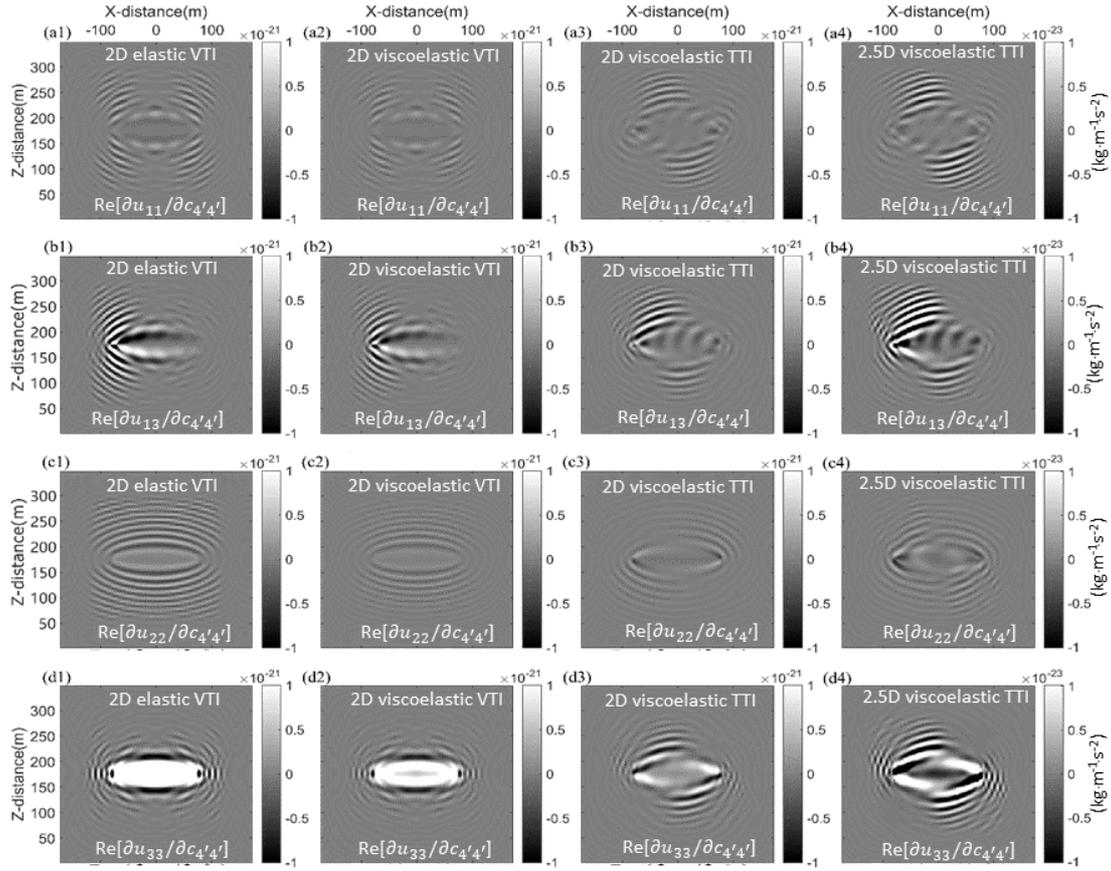

Figure 7. Numerical results of $\mathrm{Re}[\partial u_{\hat{s}\hat{g}}/\partial c_{4'4'}]$ in four homogeneous media: (1) 2-D elastic VTI, (2) 2-D viscoelastic VTI, (3) 2-D viscoelastic TTI ($\theta=30°$) and (4) 2.5-D viscoelastic TTI ($\theta=30°$).

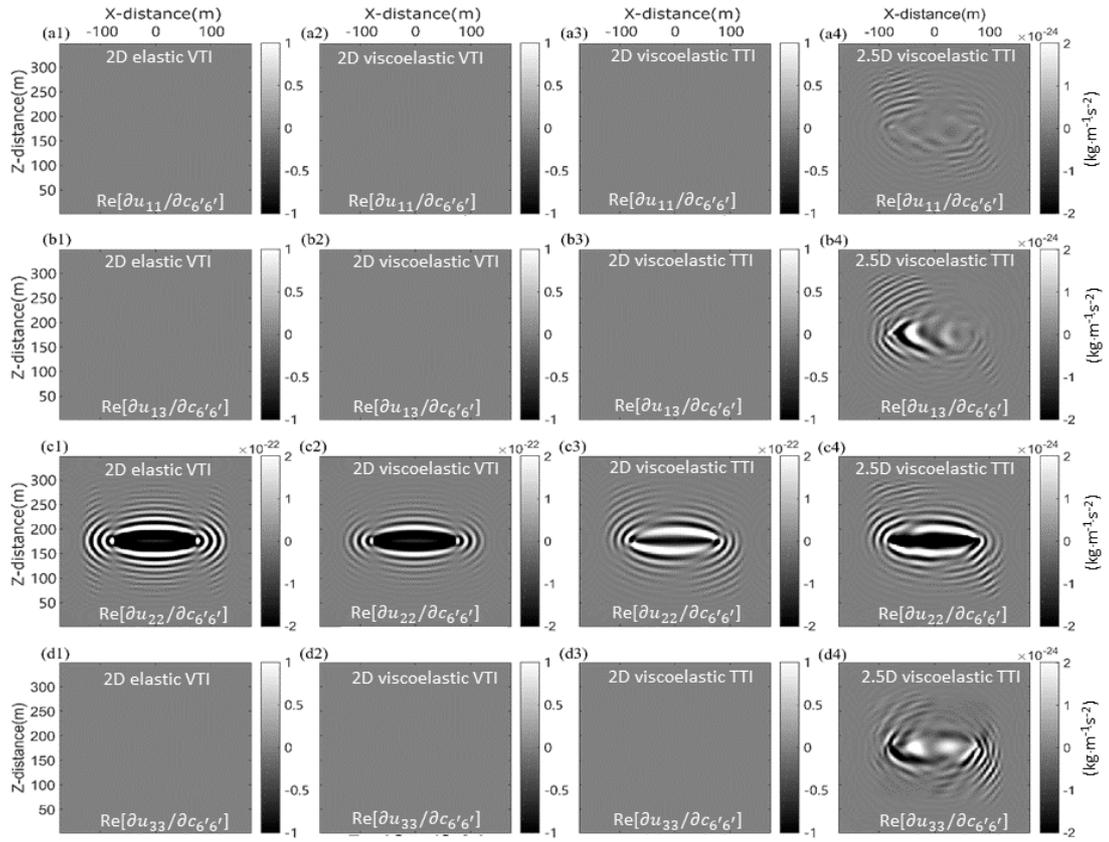

Figure 8. Numerical results of $\mathrm{Re}[\partial u_{\hat{s}\hat{g}}/\partial c_{6'6'}]$ in four homogeneous media: (1) 2-D elastic VTI, (2) 2-D viscoelastic VTI, (3) 2-D viscoelastic TTI ($\theta=30°$) and (4) 2.5-D viscoelastic TTI ($\theta=30°$).

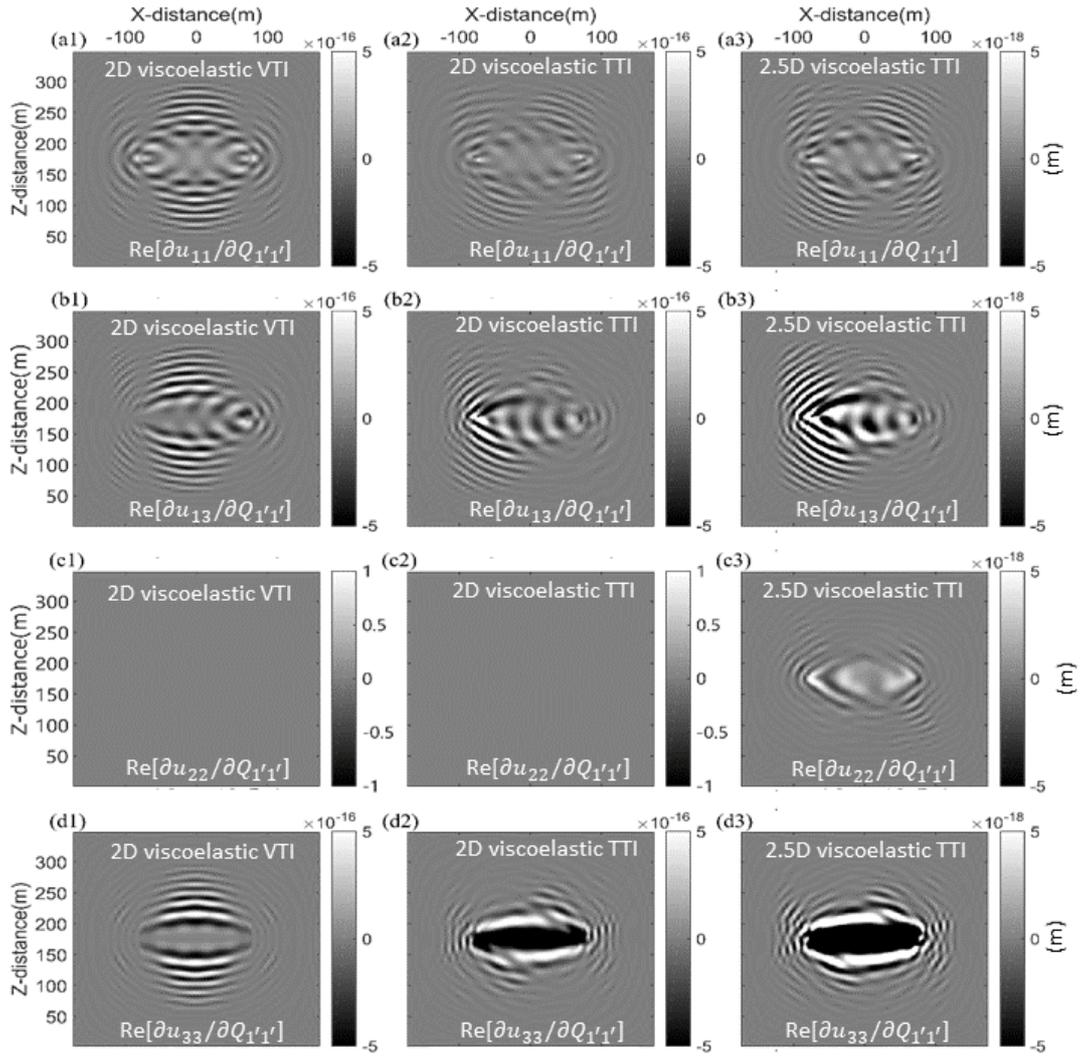

Figure 9. Numerical results of $\text{Re}[\partial u_{\hat{s}\hat{g}}/\partial Q_{1'1'}]$ in three homogeneously viscoelastic media: (1) 2-D VTI, (2) 2-D TTI ($\theta=30°$) and (3) 2.5-D TTI ($\theta=30°$).

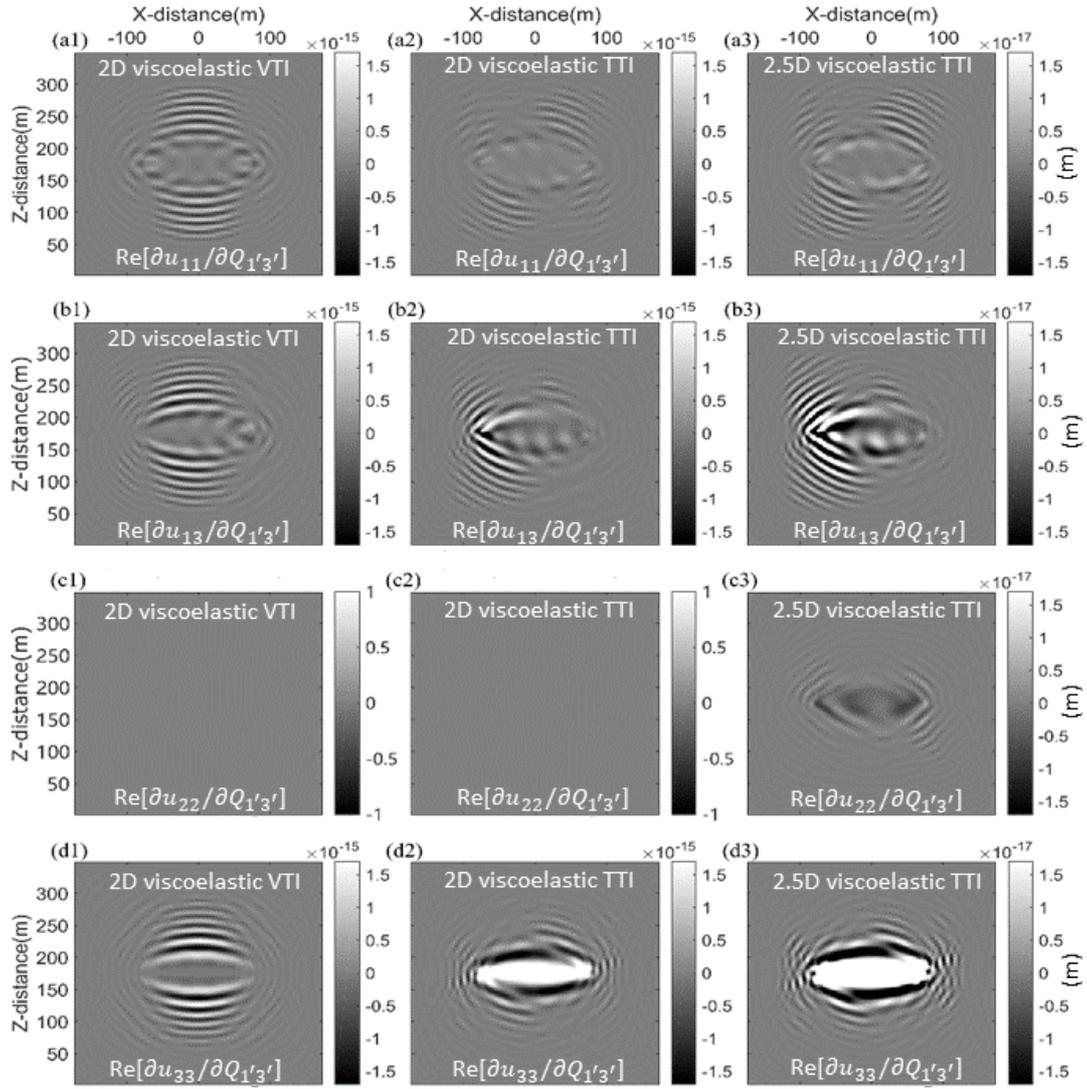

Figure 10. Numerical results of $\text{Re}[\partial u_{\hat{s}\hat{g}}/\partial Q_{1'3'}]$ in three homogeneously viscoelastic media: (1) 2-D VTI, (2) 2-D TTI ($\theta=30°$) and (3) 2.5-D TTI ($\theta=30°$).

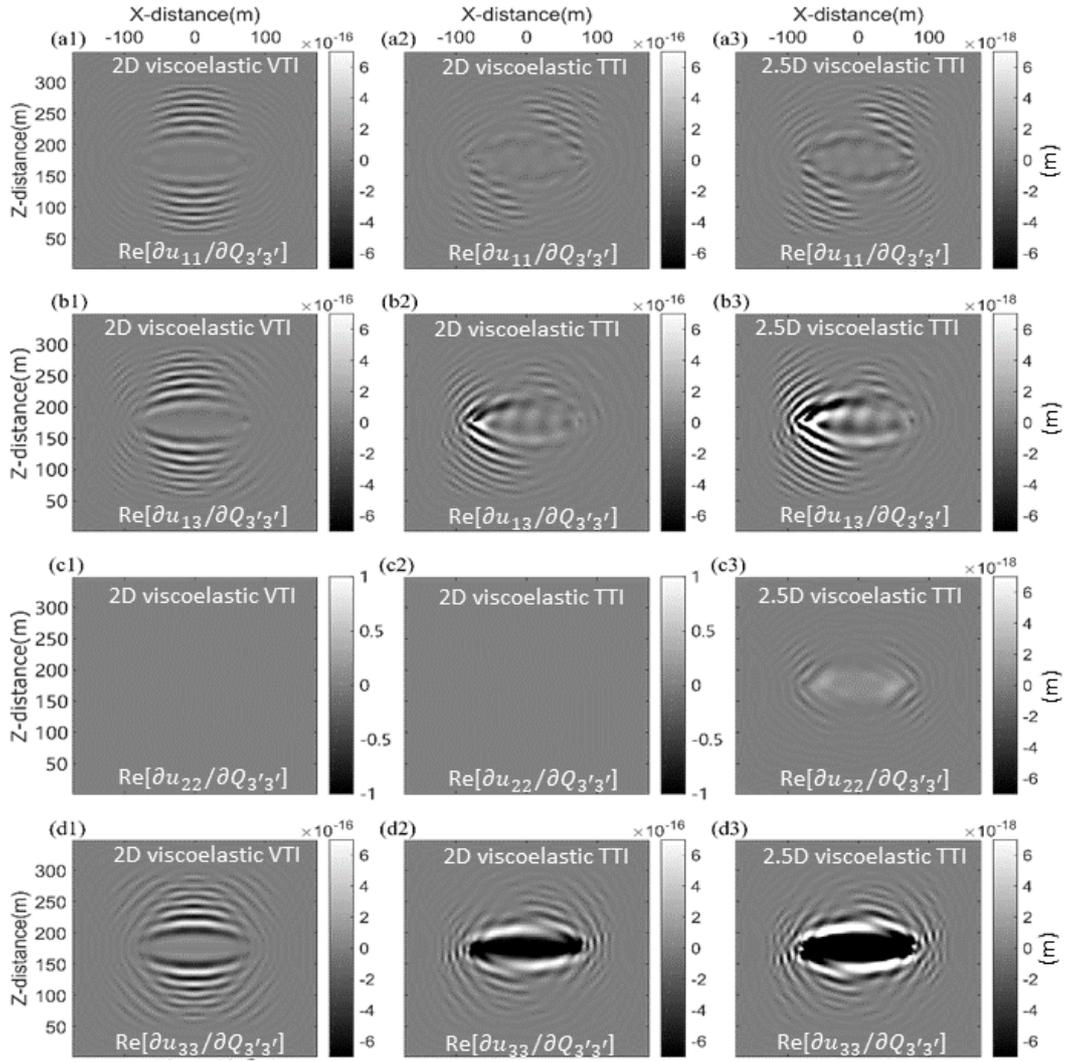

Figure 11. Numerical results of $\text{Re}\left[\partial u_{\hat{s}\hat{g}}/\partial Q_{3'3'}\right]$ in three homogeneously viscoelastic models: (1) 2-D VTI, (2) 2-D TTI ($\theta=30°$) and (3) 2.5-D TTI ($\theta=30°$).

Figure 12. Numerical results of $\text{Re}[\partial u_{\hat{s}\hat{g}}/\partial Q_{4'4'}]$ in three homogeneously viscoelastic models: (1) 2-D VTI, (2) 2-D TTI ($\theta=30°$) and (3) 2.5-D TTI ($\theta=30°$).

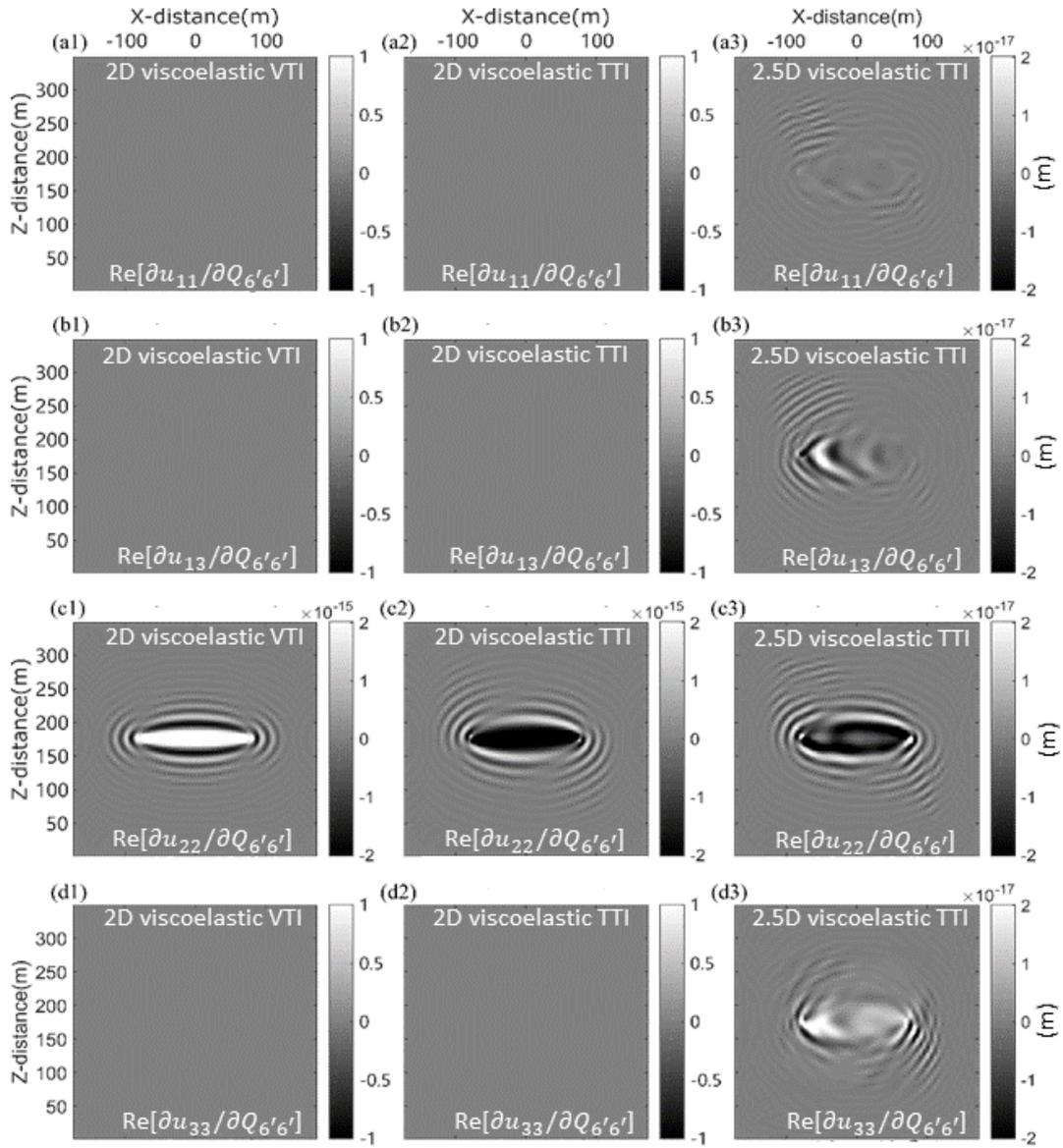

Figure 13. Numerical results of $\text{Re}[\partial u_{\hat{s}\hat{g}}/\partial Q_{6'6'}]$ in three homogeneously viscoelastic models: (1) 2-D VTI, (2) 2-D TTI ($\theta=30°$) and (3) 2.5-D TTI ($\theta=30°$).

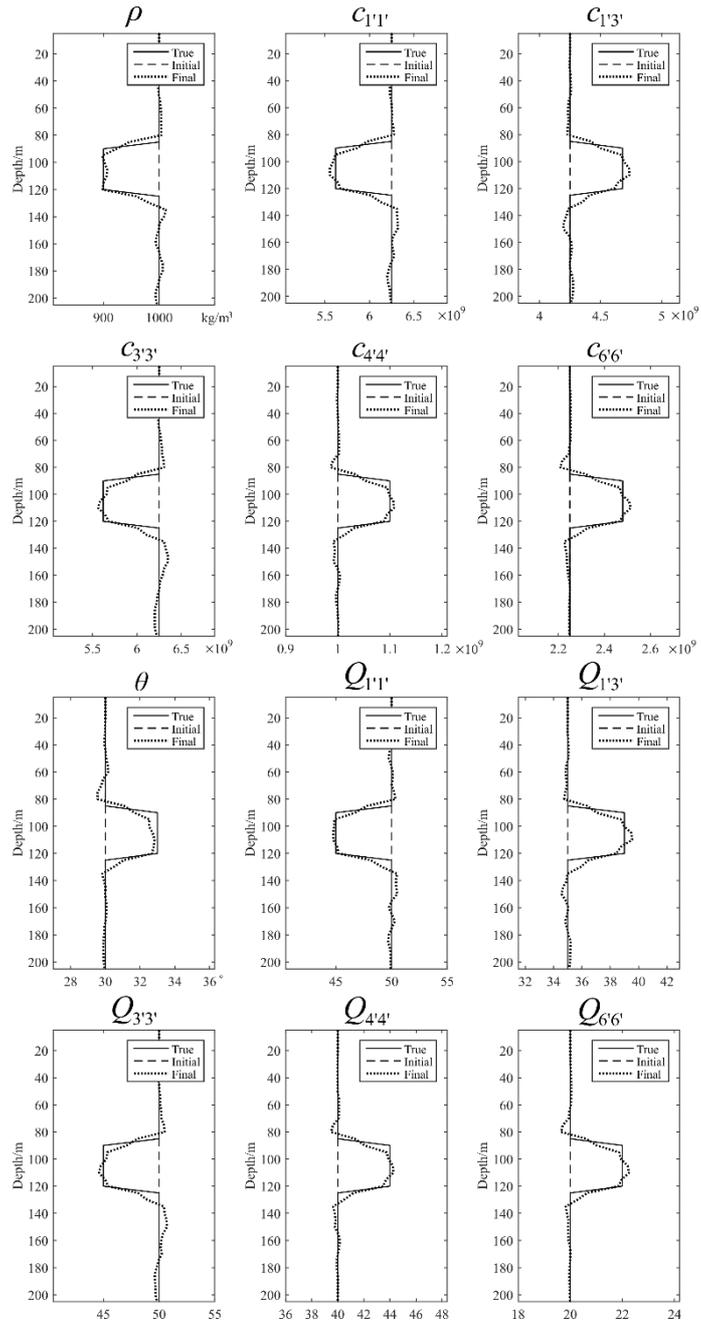

Figure 14. Inverted twelve model parameters by the frequency-domain FWI for a simple viscoelastic TTI medium: the solid lines are the true models, the dashed lines are the initial models, and the dotted lines are the final inversion results.

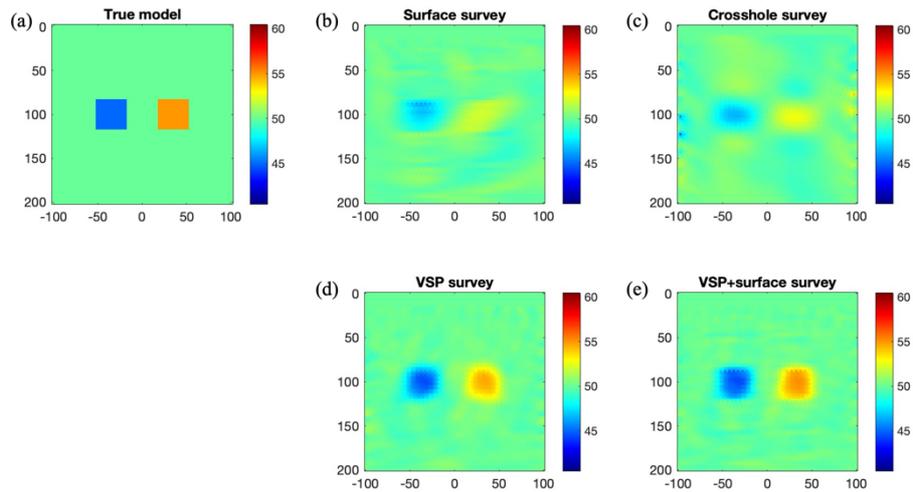

Figure 15. Inverted $Q_{1'1'}$ model for different surveying geometries. (a) the true model, (b) surface survey: 10 sources and 10 receivers on surface, (c) crosshole survey: 10 sources in the left borehole and 10 receivers in the right borehole, (d) VSP survey: 10 sources on the surface and 20 receivers in the left and right boreholes, (e) VSP+surface survey: 10 sources on the surface and 30 receivers on the surface and two boreholes.